\documentclass{aa}  
\usepackage{graphicx}
\usepackage{txfonts}
\begin{document}

\title{Radiative transfer and the energy equation \\ in SPH simulations of star formation}

\author{D. Stamatellos\inst{1}
        \and
        A.~P. Whitworth\inst{1}
        \and 
        T.~Bisbas\inst{1}	
        \and
        S.~Goodwin\inst{2}
        }

\offprints{D. Stamatellos \\ \email{D.Stamatellos@astro.cf.ac.uk}}

\institute{School of Physics \& Astronomy, Cardiff University, 5 The Parade, Cardiff, CF24 3AA, Wales, UK
\and Department of Physics and Astronomy, The University of Sheffield, 
Hicks Building, Hounsfield Road, Sheffield S3 7RH, UK}

\date{Received February, 2007; accepted ...}

% \abstract{}{}{}{}{} 
% 5 {} token are mandatory
  
\abstract
{}
{We introduce and test a new and highly efficient method for treating the thermal and radiative effects influencing the energy equation in SPH simulations of star formation.}
{The method uses the density, temperature and gravitational potential of each particle to estimate a mean optical depth, which then regulates the particle's heating and cooling. The method captures -- at minimal computational cost -- the effects of (i) the rotational and vibrational degrees of freedom of H$_2$; (ii) H$_{_2}$ dissociation and H$^{\rm o}$ ionisation; (iii) opacity changes due to ice mantle melting, sublimation of dust, molecular lines, H$^-$, bound-free and free-free processes and electron scattering; (iv) external irradiation; and (v) thermal inertia.}
{We use the new method to simulate the collapse of a $1\,{\rm M}_\odot$ cloud of initially  uniform density and temperature. At first, the collapse proceeds almost isothermally ($T\propto\rho^{0.08}$; cf. Larson 2005). The cloud starts heating fast when the optical depth to the cloud centre reaches unity ($\rho_{_{\rm C}}\sim 7\times10^{-13}~{\rm g\ cm^{-3}}$). The first core forms at $\rho_{_{\rm C}}\sim 4\times10^{-9}~{\rm g\ cm^{-3}}$ and steadily increases in mass. When the temperature at the centre reaches $T_{_{\rm C}}\sim 2000\,{\rm K}$, molecular hydrogen starts to dissociate and the second collapse begins, leading to the formation of the second (protostellar) core. The results mimic closely the detailed calculations of Masunaga \& Inutsuka (2000). We also simulate (i) the collapse of a $1.2\,{\rm M}_\odot$ cloud, which initially has uniform density and temperature, (ii) the collapse of a $1.2\,{\rm M}_\odot$ rotating cloud, with an $m=2$ density perturbation and uniform initial temperature, and (iii) the smoothing of temperature fluctuations in a static, uniform density sphere. In all these tests the new algorithm reproduces the results of previous authors and/or known analytic solutions. The computational cost is comparable to a standard SPH simulation with a simple barotropic equation of state. The method is easy to implement, can be applied to both particle- and grid-based codes, and handles optical depths $0< \tau\la 10^{11}$.
}
{}


\maketitle

%%%%%%%%%%%%%% SECTION
\section{Introduction}
%%%%%%%%%%%%%%%%%%%%%%

Smoothed Particle Hydrodynamics (SPH) (Lucy 1977; Gingold \& Monaghan 1977) is a Lagrangian method which invokes a large ensemble of particles to describe a fluid, by assigning properties such as mass, $m_i$, position, ${\bf r}_i$, and velocity, ${\bf v}_i$, to each particle, $i$. Intensive thermodynamic variables like density and pressure (and their derivatives) are estimated using local averages (for reviews see Benz 1990, 1991; Monaghan 1992, 2005). 

Radiative transfer (RT) has only recently been included in SPH codes (Oxley \& Woolfson 2003; Whitehouse \& Bate 2004, 2006; Whitehouse et al. 2005; Viau et al. 2005; Mayer et al. 2007). These SPH-RT codes use different simplifying assumptions in order to by-pass the full treatment of multi-frequency radiative transfer in 3 dimensions (a task which is not possible with current computing resources), but they still tend to be computationally expensive. Indeed, even the treatment of full 3D radiative transfer on a single snapshot during the evolution of a simulation is computationally quite expensive (Stamatellos \& Whitworth 2005; Stamatellos et al. 2005).

More often, in  SPH simulations of star formation, it is standard practice to use a barotropic equation of state, i.e. to put $P = P(\rho)$ (e.g. Bonnell 1994; Whitworth et al. 1995; Bate 1998). The form of $P(\rho)$ is chosen to mimic the thermodynamics of star forming gas, as revealed by computations of the spherically symmetric collapse of a single, isolated protostar (e.g. Boss \& Myhill 1992; Masunaga \& Inutsuka 2000). 

This is not an ideal situation. (a) A barotropic equation of state is unable to account for the fact that the thermal history of a protostar depends sensitively on its environment, geometry and mass; for example, low-mass protostars remain optically thin to their cooling radiation to higher densities than high-mass ones. Thus, the evolution of the density and temperature cannot be approximated by a single barotropic equation for every system. Even for the same system, the density and temperature evolution away from the centre of the cloud does not follow the corresponding evolution at the centre of the cloud (Whitehouse \& Bate 2006). (b) A barotropic equation of state is unable to capture thermal inertia effects (i.e. situations where the evolution is controlled by the thermal timescale, rather than the dynamical one). Such effects appear to be critical at the stage when fragmentation occurs (e.g. Boss et al. 2000). 

One of the ultimate goals of star formation simulations is to track the thermal history of star-forming gas. Strictly speaking, this requires a computational method which can treat properly, in 3 dimensions, the time-dependent radiation transport which controls the energy equation. However, this is computationally very expensive, significantly more expensive than the hydrodynamics. We have therefore developed a new algorithm which enables us to distinguish the thermal behaviours of protostars of different mass, in different environments, with different metallicities, {\it and} to capture thermal inertia effects, {\it without} treating in detail the associated radiation transport. The method uses the density, temperature and gravitational potential of each SPH particle (which are all calculated using the standard SPH formalism) to estimate a characteristic optical depth for each particle. This optical depth then regulates how each particle heats and cools.

The paper is organized as follows. In Section~\ref{sec:method}, we present the new algorithm we have developed to treat the energy equation within SPH, also describing the aspects of SPH that are required for a complete picture of the new method. In Section~\ref{sec:dustgas} we outline the properties we adopt for the gas and dust in a star-forming cloud, i.e. the composition, energy equation, equation of state, and opacity. In Section~\ref{sec:mmi}, we  present a simulation of the collapse of a $1\,{\rm M}_\odot$ molecular cloud; we describe in detail the different stages of the collapse, and compare our results with those of Masunaga \& Inutsuka (2000). In Section~\ref{sec:tests}, three additional tests are presented: (i) the collapse of a $1.2\,{\rm M}_\odot$ molecular cloud (Boss \& Myhill 1992; Whitehouse \& Bate 2006), (ii) the collapse of a rotating $1.2\,{\rm M}_\odot$ molecular cloud, with an $m=2$ density perturbation (Boss \& Bodenheimer 1979; Whitehouse \& Bate 2006), and (iii) the smoothing of temperature fluctuations in a static, uniform-density sphere (Spiegel 1957). Finally, in Section~\ref{sec:sum}, we summarise the method and the tests performed, and discuss the applicability of the new algorithm to simulations of astrophysical systems.

%%%%%%%%%%%%%%%%%%%%%%%%%%%%%% SECTION
\section{The method}\label{sec:method}
%%%%%%%%%%%%%%%%%%%%%%%%%%%%%%%%%%%%%%

The key to the new method is to use an SPH particle's density, $\rho_i$, temperature, $T_i$, and gravitational potential, $\psi_i$, to estimate a mean optical depth, $\bar{\tau}_i$, for the SPH particle. This mean optical depth then regulates the SPH particle's radiative heating and cooling; in other words, it determines the extent to which the SPH particle is shielded from external radiation, and the extent to which the SPH particle's cooling radiation is trapped. (The gravitational potential is used here, purely because gravity is the only particle parameter which is already calculated by the SPH code but is not a local function of state. Therefore it should, in some very general sense, represent the larger-scale environment surrounding the SPH particle.)

Specifically, each SPH particle is treated as if it were embedded in a spherically-symmetric pseudo-cloud (its personal pseudo-cloud). The density and temperature profiles of the pseudo-cloud are modelled with a polytrope of index $n=2$, but the pseudo-cloud is not assumed to be in hydrostatic balance. (We will show later that the choice of $n$ is not critical.)

The position of the SPH particle within its pseudo-cloud is not specified; instead we take a mass-weighted average over all possible positions (see Figs. 1 \& 2). For any given position of the SPH particle within the pseudo-cloud, the central density, $\rho_{_{\rm C}}$, and scale-length, $R_{_{\rm O}}$, are chosen to reproduce the density and gravitational potential at the position of the SPH particle. Similarly, the pseudo-cloud's central temperature, $T_{_{\rm C}}$ is chosen to match the temperature at the position of the SPH particle. (Because the pseudo-cloud is not necessarily in hydrostatic equilibrium, we cannot -- in general -- write $T_{_{\rm C}}=4\pi G\bar{m}\rho_{_{\rm C}}R^2_{_{\rm O}}/(n+1)k_{_{\rm B}}$, where $\bar{m}$ is the mean gas-particle mass and $k_{_{\rm B}}$ is Boltzmann's constant.)

The optical depth, $\tau_i$, is then calculated by integrating out along a radial line from the given position to the edge of the pseudo-cloud, i.e. through the cooler and more diffuse outer parts of the pseudo-cloud. In this way, $\tau_i$ samples the different opacity regimes that are likely to surround the SPH particle. This accounts for the fact that, even if the opacity at the position of the SPH particle is low, its cooling radiation may be trapped by cooler more opaque material in the surroundings.

Finally, $\bar{\tau}_i$ is obtained by taking a mass-weighted average over all possible positions within the pseudo-cloud.

\subsection{Basic SPH equations}
%%%%%%%%%%%%%%%%%%%%%%%%%%%%%%%%

We use the {\sc dragon} SPH code (Goodwin et al. 2004a, b). {\sc dragon} uses variable smoothing lengths, $h_i$, adjusted so that the number of neighbours is exactly ${\cal N}_{_{\rm NEIB}}=50$; it is important to have a constant number of neighbours to minimise numerical diffusion  (Attwood et al. 2007). An octal tree is used to collate neighbour lists and calculate gravitational accelerations, which are kernel-softened using particle smoothing lengths. Standard artificial viscosity is invoked in converging regions, and multiple particle time-steps are used.

The density at the position of SPH particle $i$ is given by a sum over its neighbours, $j$,
\begin{eqnarray}
\rho_i&=&\sum_j\left\{\frac{m_j}{h_{ij}^3}\,W\left(\frac{{r}_{ij}}{{h}_{ij}}\right)\right\}\,,
\end{eqnarray}
where $m_j$ is the mass of particle $j$, $h_{ij}=(h_i+h_j)/2$, and $W(s)$ is the dimensionless smoothing kernel.

The equation of motion for SPH particle $i$ is
\begin{eqnarray}\label{momentum}
\frac{d{\bf v}_i}{dt}&=&\sum_{j\ne i}\!\left\{m_j\left(\frac{P_i}{\rho_i^2}\!+\!\frac{P_j}{\rho_j^2}\!+\!\Pi_{ij}\right)\frac{{\bf r}_{ij}}{h_{ij}^4r_{ij}}W^\prime\!\left( \frac{{ r}_{ij}}{{h}_{ij}} \right)\right\}\!+\!\left.\frac{d{\bf v}_i}{dt}\right|_{_{\rm \,GRAV}}\!.
\end{eqnarray}
Here, $P_i$ and $P_j$ are the pressures at the positions of particles $i$ and $j$ respectively; $\,{\bf r}_{ij}\equiv{\bf r}_i-{\bf r}_j\,$; $\,r_{ij}\equiv|{\bf r}_{ij}|\,$; and $\,W'(s)\equiv dW/ds$. Artificial viscosity is represented by the term
\begin{eqnarray}
\Pi_{ij}&=&\frac{(-\alpha c_{ij}\mu_{ij}+\beta\mu_{ij}^2)}{\rho_{ij}}\,,
\end{eqnarray}
with $\,\alpha=1\,$, $\,\beta=2\,$, $\,c_{ij}\equiv(c_i+c_j)/2\;\,$ (where $c_i=(P_i/\rho_i)^{1/2}$ is the isothermal sound speed of particle $i$), $\,\rho_{ij}\equiv(\rho_i+\rho_j)/2\,$, 
\begin{eqnarray}
\mu_{ij}&=&\left\{\begin{array}{ll}
\frac{{h}_{ij}\,{\bf v}_{ij}\cdot{\bf r}_{ij}}{(r_{ij}^2+0.1 h_{ij}^2)}\,,\hspace{0.5cm} & {\rm if}\;{\bf v}_{ij}\cdot{\bf r}_{ij}<0\,, \\
0\,, & {\rm if}\;{\bf v}_{ij}\cdot{\bf r}_{ij}\geq0\,,
\end{array}\right.
\end{eqnarray}
and ${\bf v}_{ij}\equiv{\bf v}_i-{\bf v}_j$.

The second term on the righthand side of Eqn. (\ref{momentum}) is the gravitational acceleration experienced by particle $i$, given by
\begin{eqnarray}\label{EQN:GRAVACC}
\left.\frac{d{\bf v}_i}{dt}\right|_{_{\rm \,GRAV}}&=&-\,\sum_{j\ne i}\left\{\frac{Gm_j{\bf r}_{ij}}{r^3_{ij}}W^\star\!\left(\frac{r_{ij}}{h_{ij}}\right)\right\}\,,
\end{eqnarray}
where
\begin{eqnarray}\label{EQN:Wstar}
W^\star(s)&=&\int_{s'=0}^{s'=s}\,W(s')\,4\,\pi\,s'^2\,ds'\,.
\end{eqnarray}
Similarly the gravitational potential at the position of particle $i$ is given by
\begin{eqnarray}\label{EQN:GRAVPOT}
\psi_i=-\,G\,\sum_{j\ne i}\left\{\frac{m_j}{{r}_{ij}}\,W^{\star\star}\left(\frac{{r}_{ij}}{{h}_{ij}}\right)\right\}\,,
\end{eqnarray}
where
\begin{eqnarray}\label{EQN:Wstarstar}
W^{\star\star}(s)=W^\star(s)\,+\,s\int_{s'=s}^{s'=\infty}W(s')\,4\,\pi\,s'\,ds'\,.
\end{eqnarray}
In Eqns. (\ref{EQN:GRAVACC}) and (\ref{EQN:GRAVPOT}), the sums are over all particles except $i$; the terms $W^\star(s)$ in Eqn. (\ref{EQN:Wstar}) and $W^{\star\star}(s)$ in Eqn. (\ref{EQN:Wstarstar}) represent kernel softening. In practice, the calculation of these gravitational terms is rendered more efficient by using a tree structure to identify distant clusters of particles whose effect can be treated collectively with a multipole expansion; this reduces an ${\cal N}^2$ process to an ${\cal N}\ell n[{\cal N}]$ process, where ${\cal N}$ is the total number of SPH particles.

The energy equation for particle $i$ is
\begin{eqnarray}\label{EQN:ENERGY}
\frac{du_i}{dt}=\frac{1}{2}\!\sum_j\!\left\{m_j\!\left(\frac{P_i}{\rho_i^2}\!+\!\frac{P_j}{\rho_j^2}\!+\!\Pi_{ij}\right)\!\frac{{\bf v}_{ij}\cdot{\bf r}_{ij}}{h_{ij}^4\,r_{ij}}\!W'\left(\frac{r_{ij}}{h_{ij}}\right)\right\}\!+\!\left.\frac{du_i}{dt}\right|_{_{\rm RAD}}\!.
\end{eqnarray}
Here $u_i$ is the internal energy {\it per unit mass}. The first term on the righthand side represents compressional and viscous heating. The second term on the righthand side is the net radiative heating rate; this paper is primarily concerned with the evaluation of this term.

%%%%%%%%%%%%%%%%%%%%%%%%%%%%%%%%%%%%%%%%%%%%%%%%%%%%% FIGURE
\begin{figure}
\centerline{\includegraphics[height=5.cm]{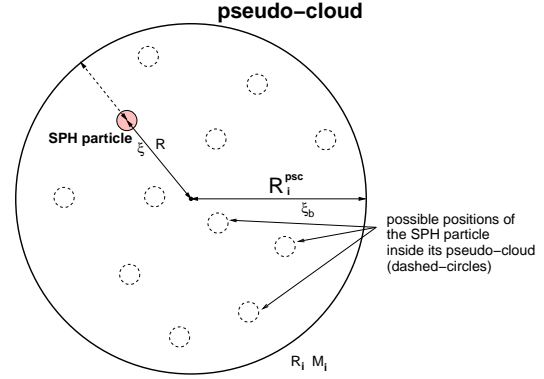}}
\caption{Schematic representation of the pseudo-cloud around an SPH particle.The location of the SPH particle inside its pseudo-cloud is not specified.}
\label{fig:pscloud}
\end{figure}
%%%%%%%%%%%%%%%%%%%%%%%%%%%%%%%%%%%%%%%%%%%%%%%%%%%%%%%%%%%%
%%%%%%%%%%%%%%%%%%%%%%%%%%%%%%%%%%%%%%%%%%%%%%%%%%% FIGURE
\begin{figure}[b]
\centerline{\includegraphics[height=9.cm]{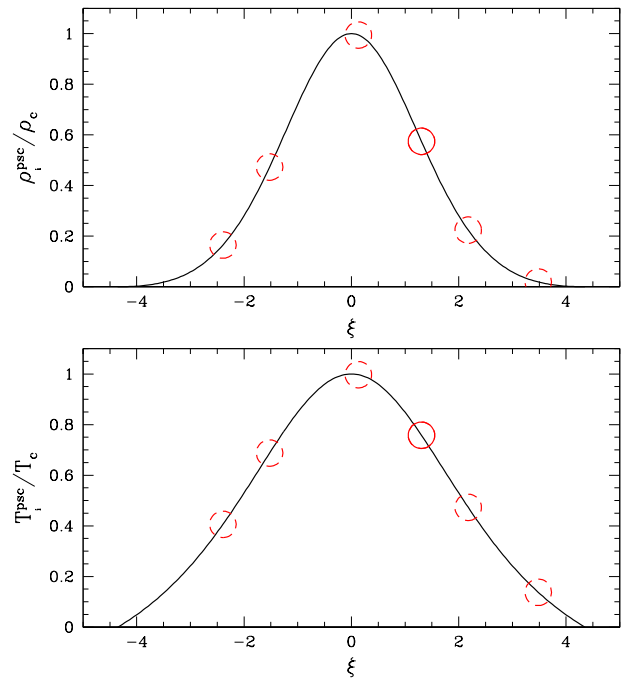}}
\caption{Density and temperature profiles for a polytropic pseudo-cloud with $n=2$. The SPH particle could be located anywhere in the cloud (solid and dashed line circles).}
\label{fig:polytrope}
\end{figure}
%%%%%%%%%%%%%%%%%%%%%%%%%%%%%%%%%%%%%%%%%%%%%%%%%%%%%%%%%%
\subsection{Calibrating the pseudo-cloud}\label{SEC:CALIBR}
%%%%%%%%%%%%%%%%%%%%%%%%%%%%%%%%%%%%%%%%%%%%%%%%%%%%%%%%%%%

Suppose that SPH particle $i$ is embedded at radius $R=\xi R_{_0}$, in a pseudo-cloud with central density $\rho_{_{\rm C}}$, scale-length $R_{_{\rm O}}$, and polytropic index $n$ (see Figs. 1 \& 2). $\;\,\xi\,$ is thus a dimensionless radius, and $\rho_{_{\rm C}}$ and $R_{_{\rm O}}$ are chosen so as to reproduce -- at this radius -- the actual density and gravitational potential of the SPH particle, i.e.
\begin{eqnarray}
\rho_{_{\rm C}}\theta^n(\xi)&=&\rho_i\,,\\
-\,4\pi G\rho_{_{\rm C}}R_{_{\rm O}}^2\phi(\xi)&=&\psi_i\,.
\end{eqnarray}
Here $\theta(\xi)$ is the Lane-Emden Function for index $n$ (Chandrasekhar 1939), 
\begin{eqnarray}
\phi(\xi)&=&-\,\xi_{_{\rm B}}\,\frac{d\theta}{d\xi}\left(\xi_{_{\rm B}}\right)\,+\,\theta(\xi)\,,
\end{eqnarray}
and $\xi_{_{\rm B}}$ is the dimensionless boundary of the polytrope (i.e. the argument of the smallest zero of $\theta(\xi)$)

If we fix $n$ (and hence the forms of $\theta(\xi)$ and $\phi(\xi)$), and we pick an arbitrary value for $\xi$ (modulo that it must be within the pseudo-cloud, i.e. $\xi<\xi_{_{\rm B}}$), then we obtain
\begin{eqnarray}
\rho_{_{\rm C}}&=&\rho_i\,\theta^{-n}(\xi)\,,\\
R_{_{\rm O}}&=&\left[\frac{-\,\psi_i\,\theta^n(\xi)}{4\,\pi\,G\,\rho_i\,\phi(\xi)}\right]^{1/2}\,.
\end{eqnarray}

In an analogous manner we chose the central temperature of the pseudo-cloud so as to reproduce -- at radius $R=\xi R_{_{\rm O}}$ -- the actual temperature of the SPH particle (see Fig. 2),
\begin{eqnarray}
T_{_{\rm C}}\,\theta(\xi)&=&T_i\,,\\
T_{_{\rm C}}&=&T_i\,\theta^{-1}(\xi)\,.
\end{eqnarray}

The column-density on a radial line from this radius to the boundary of the pseudo-cloud is then given by
\begin{eqnarray}\nonumber
\Sigma_i(\xi)&=&\int_{\xi'=\xi}^{\xi'=\xi_{_{\rm B}}}\,\rho_{_{\rm C}}\theta^n(\xi')\,R_{_{\rm O}}d\xi'\\
&=&\left[\frac{-\,\psi_i\,\rho_i}{4\,\pi\,G\,\phi(\xi)\,\theta^n(\xi)}\right]^{1/2}\,\int_{\xi'=\xi}^{\xi'=\xi_{_{\rm B}}}\theta^n(\xi')\,d\xi'\,.
\end{eqnarray}
To obtain the pseudo-mean column-density, we take a mass-weighted average of $\Sigma_i(\xi)$ over all possible dimensionless radii, $\xi$, i.e.
\begin{eqnarray}\nonumber
\bar{\Sigma}_i&=&\left[-\,\xi_{_{\rm B}}^2\,\frac{d\theta}{d\xi}(\xi_{_{\rm B}})\right]^{-1}\,\int_{\xi=0}^{\xi=\xi_{_{\rm B}}}\;\;\Sigma_i(\xi)\;\;\,\theta^n(\xi)\,\xi^2d\xi\\\label{EQN:SIGMABAR}
&=&\zeta_n\,\left[\frac{-\,\psi_i\,\rho_i}{4\,\pi\,G}\right]^{1/2}\,,
\end{eqnarray}
where $\left[-\,\xi_{_{\rm B}}^2\,\frac{d\theta}{d\xi}(\xi_{_{\rm B}})\right]$ is the total dimensionless mass of the polytrope, $\theta^n(\xi)\,\xi^2d\xi$ is the dimensionless mass element between $\xi$ and $\xi+d\xi$, and 
\begin{eqnarray}
\zeta_n\!&\!=\!&\!\left[-\,\!\xi_{_{\rm B}}^2\,\!\frac{d\theta}{d\xi}(\xi_{_{\rm B}})\right]^{-1}\!\!\int_{\xi=0}^{\xi=\xi_{_{\rm B}}}\!\!\int_{\xi'=\xi}^{\xi'=\xi_{_{\rm B}}}\!\theta^n(\xi')\,\!d\xi'\;\left[\frac{\theta^n(\xi)}{\phi(\xi)}\right]^{1/2}\!\xi^2\,\!d\xi.
\end{eqnarray}

As an indication of how insensitive the results are to the choice of $n$, we note that $\zeta_{_1}=0.376,\;\zeta_{_{1.5}}=0.372,\;\zeta_{_2}=0.368,\;\zeta_{_{2.5}}=0.364,\;{\rm and}\;\zeta_{_3}=0.360\,.\,$ Since for protostars which are close to equilibrium -- for example those undergoing quasistatic (i.e. Kelvin-Helmholtz) contraction -- the polytropic exponent is likely to fall in the range $4/3\;{\rm to}\;5/3$, we adopt $n=2$ (corresponding to a polytropic exponent of 3/2).

We calculate the pseudo-mean optical depth in the same way as the pseudo-mean column-density. If the Rosseland-mean opacity is a function of density and temperature only, $\kappa_{_{\rm R}}(\rho,T)$, the radial optical depth from radius $R=\xi R_{_{\rm O}}$ to the boundary of the pseudo-cloud is
\begin{eqnarray}\nonumber
\tau_i(\xi)&=&\int_{\xi'=\xi}^{\xi'=\xi_{_{\rm B}}}\,\kappa_{_{\rm R}}\left(\rho_{_{\rm C}}\theta^n(\xi'),T_{_{\rm C}}\theta(\xi')\right)\,\rho_{_{\rm C}}\theta^n(\xi')\,R_{_{\rm O}}d\xi'\\\nonumber
&=&\left[\frac{-\,\psi_i\,\rho_i\,\theta^n(\xi)}{4\,\pi\,G\,\phi(\xi)}\right]^{1/2}\,\times\\
&&\;\int_{\xi'=\xi}^{\xi'=\xi_{_{\rm B}}}\,\kappa\left(\rho_i\left[\frac{\theta(\xi')}{\theta(\xi)}\right]^n,T_i\left[\frac{\theta(\xi')}{\theta(\xi)}\right]\right)\;\left[\frac{\theta(\xi')}{\theta(\xi)}\right]^n\,d\xi'\,,
\end{eqnarray}
and the mass-weighted pseudo-mean optical depth is
\begin{eqnarray}\nonumber
\bar{\tau}_i&=&\left[-\,\xi_{_{\rm B}}^2\,\frac{d\theta}{d\xi}(\xi_{_{\rm B}})\right]^{-1}\,\left[\frac{-\,\psi_i\,\rho_i}{4\,\pi\,G}\right]^{1/2}\int_{\xi=0}^{\xi=\xi_{_{\rm B}}}\!\int_{\xi'=\xi}^{\xi'=\xi_{_{\rm B}}}\,\times\\\label{EQN:TAUBAR}
&&\,\kappa_{_{\rm R}}\!\left(\rho_i\!\left[\frac{\theta(\xi')}{\theta(\xi)}\right]^n\!,T_i\!\left[\frac{\theta(\xi')}{\theta(\xi)}\right]\right)\,\!\theta^n(\xi')\,\!d\xi'\!\left[\frac{\theta^n(\xi)}{\phi(\xi)}\right]^{1/2}\,\!\xi^2\,\!d\xi\,\!.
\end{eqnarray}

At first sight it might appear that the double integral in Eqn. (\ref{EQN:TAUBAR}) will have to be evaluated on-the-fly for every SPH particle, at every time-step. In fact, we can define a pseudo-mean mass opacity
\begin{eqnarray}
\bar{\kappa}_i&=&\frac{\bar{\tau}_i}{\bar{\Sigma}_i}\,,
\end{eqnarray}
which -- once $n$ has been fixed -- is simply a function of $\rho_i$ and $T_i$. It can therefore be evaluated in advance, once and for all time, and stored in a dense look-up table for subsequent reference and interpolation. For the point $(\rho,T)$ in the table,
\begin{eqnarray}\nonumber
\bar{\kappa}_{_{\rm R}}(\rho,T)\!&\!=\!&\!\left[-\,\zeta_n\,\xi_{_{\rm B}}^2\,\frac{d\theta}{d\xi}(\xi_{_{\rm B}})\right]^{-1}\,\int_{\xi=0}^{\xi=\xi_{_{\rm B}}}\int_{\xi'=\xi}^{\xi'=\xi_{_{\rm B}}}\,\times\\\label{EQN:MEANOPACITY}
&&\kappa_{_{\rm R}}\left(\rho\left[\frac{\theta(\xi')}{\theta(\xi)}\right]^n,T\left[\frac{\theta(\xi')}{\theta(\xi)}\right]\right)\left[\frac{\theta^{\,n+2}(\xi)}{\phi(\xi)}\right]^{1/2}\!d\xi'\,\xi^2d\xi.
\end{eqnarray}

The physical interpretation of this pseudo-mean opacity, $\bar{\kappa}_{_{\rm R}}(\rho,T)$, is fundamental to the method. Although formally $\bar{\kappa}_{_{\rm R}}(\rho,T)$ only depends on the local density and temperature, when it is multiplied by the pseudo-mean column-density, $\bar{\Sigma}_i$, it gives a pseudo-mean optical depth, $\bar{\tau}_i$, which allows for the fact that radiation absorbed or emitted by particle $i$ has to pass through surrounding material which will in general have different density and temperature, and hence different opacity. For example, an SPH particle whose local Rosseland-mean opacity, $\kappa_{_{\rm R}}(\rho,T)$, is low because its density and temperature fall in the opacity gap, will have a larger pseudo-mean opacity, $\bar{\kappa}_{_{\rm R}}(\rho,T)$; this simply reflects the fact that this SPH particle may still be well insulated by cooler material in its surroundings which has much higher opacity because it contains dust.

\subsection{Radiative heating and cooling}
%%%%%%%%%%%%%%%%%%%%%%%%%%%%%%%%%%%%%%%%%%

The net radiative heating for SPH particle $i$ is given by
\begin{eqnarray}
\label{EQN:RADHEAT}
\left.\frac{du_i}{dt}\right|_{_{\rm RAD}}&=&\frac{4\,\sigma_{_{\rm SB}}\,(T_{_{\rm O}}^4({\bf r}_i)-T_i^4)}{\bar{{\Sigma}}_i^2\,\bar{\kappa}_{_{\rm R}}(\rho_i,T_i)\,+\,\kappa_{_{\rm P}}^{-1}(\rho_i,T_i)}\,,
\end{eqnarray}
where $\sigma_{_{\rm SB}}$ is the Stefan-Boltzmann constant, $\bar{\kappa}_{_{\rm R}}(\rho,T)$ is the pseudo-mean opacity defined in Section \ref{SEC:CALIBR}, and $\kappa_{_{\rm P}}(\rho,T)$ is the Planck-mean opacity.

The positive term in Eqn. (\ref{EQN:RADHEAT}) -- the one involving $T_{_{\rm O}}^4({\bf r}_i)$ -- represents radiative heating due to the background radiation field with effective temperature $T_{_{\rm O}}({\bf r}_i)$. This term ensures that the SPH particle does not cool {\it radiatively} below $T_{_{\rm O}}({\bf r}_i)$. In a simulation which includes stars -- either pre-existing, or formed as an outcome of the simulation; and with luminosities $L_\star$ and positions ${\bf r}_\star$ -- we set
\begin{eqnarray}
T_{_{\rm O}}^4({\bf r})&=&\left(10\,{\rm K}\right)^4\,+\,\sum_\star\,\left\{\frac{L_\star}{16\,\pi\,\sigma_{_{\rm SB}}\,|{\bf r}-{\bf r}_\star|^2}\right\}\,.
\end{eqnarray}

The negative term in Eqn. (\ref{EQN:RADHEAT}) -- the one involving $T_i^4$ -- represents radiative cooling of SPH particle $i$. If $T_i^4\gg T_{_{\rm O}}^4({\bf r}_i)$, we can neglect the heating term and consider two limiting regimes:

(i) If $\bar{{\Sigma}}_i^2\,\bar{\kappa}_{_{\rm R}}(\rho_i,T_i)\ll\kappa_{_{\rm P}}^{-1}(\rho_i,T_i)$, we are in the optically thin cooling regime and Eqn. (\ref{EQN:RADHEAT}) approximates to
\begin{eqnarray}
\left.\frac{du_i}{dt}\right|_{_{\rm RAD}}&\simeq&-\,4\,\sigma_{_{\rm SB}}\,T_i^4\,\kappa_{_{\rm P}}(\rho_i,T_i)\,,
\end{eqnarray}
in exact agreement with the definition of the Planck-mean opacity.

(ii) If $\bar{{\Sigma}}_i^2\,\bar{\kappa}_{_{\rm R}}(\rho_i,T_i)\gg\kappa_{_{\rm P}}^{-1}(\rho_i,T_i)$, we are in the optically thick cooling regime and Eqn. (\ref{EQN:RADHEAT}) approximates to
\begin{eqnarray}\label{EQN:DIFFAP}
\left.\frac{du_i}{dt}\right|_{_{\rm RAD}}&\simeq&-\,\frac{4\,\sigma_{_{\rm SB}}\,T_i^4}{\bar{{\Sigma}}_i^2\,\bar{\kappa}_{_{\rm R}}(\rho_i,T_i)}\;=\;-\,\frac{c\,a_{_{\rm SB}}\,T_i^4}{\bar{\Sigma}_i\,\bar{\tau}_i}\,,
\end{eqnarray}
where $c$ is the speed of light, $a_{_{\rm SB}}$ is the radiant energy density constant, and we have obtained the second expression by substituting $4\sigma_{_{\rm SB}}=ca_{_{\rm SB}}$  and $\bar{{\Sigma}}_i\,\bar{\kappa}_{_{\rm R}}(\rho_i,T_i)=\bar{\tau}_i$.

To see that this is just the diffusion approximation, suppose that the pseudo-cloud has pseudo-mass $\bar{M}_i$ and pseudo-radius $\bar{R}_i\sim(3\bar{M}_i/4\pi\bar{\Sigma}_i)^{1/2}\;$. Eqn. (\ref{EQN:DIFFAP}) then reduces to 
\begin{eqnarray}
\left.\frac{du_i}{dt}\right|_{_{\rm RAD}}&\sim&\;-\,\frac{\bar{U}_{{\rm rad},\,\!i}}{\bar{M}_i\,\bar{t}_{{\rm diff},\,\!i}}\,,
\end{eqnarray}
where $\bar{U}_{{\rm rad},\,\!i}\sim4\pi\bar{R}_i^3a_{_{\rm SB}}T_i^4/3$ is the total radiant energy in the pseudo-cloud and $t_{{\rm rad},\,\!i}\sim\bar{R}_i\bar{\tau}_i/c$ is the timescale on which radiation diffuses out of the pseudo-cloud (cf. Masunaga \& Inutsuka 1999; Whitworth \& Stamatellos 2006).

\subsection{Quasi-implicit scheme}
%%%%%%%%%%%%%%%%%%%%%%%%%%%%%%%%%%

In order to avoid very short time-steps, we use the following scheme to update the internal energy, $u_i$. From SPH we know the net compressive plus viscous heating rate,
\begin{eqnarray}\label{EQN:COMPRESS}
\left.\frac{du_i}{dt}\right|_{_{\rm HYDRO}}\!&=&\frac{1}{2}\!\sum_{j}\!m_{j}\!\left\{\left( \frac{P_{i}}{\rho_{i}^2}\!+\!\frac{P_{j}}{\rho_{j}^2}\!+\!\Pi_{ij}\right)\!\frac{{\bf v}_{ij}\cdot{\bf r}_{ij}}{h_{ij}^4\,r_{ij}}W^\prime\!\left( \frac{r_{ij}}{h_{ij}}\right)\right\}\!.
\end{eqnarray}
We can therefore calculate (a) the equilibrium temperature $T_{{\rm eq},\,\!i}$ for each particle from 
\begin{eqnarray}\label{EQN:EQUILIB}
\left.\frac{du_i}{dt}\right|_{_{\rm HYDRO}}\,+\,\frac{\,4\,\sigma_{_{\rm SB}}\,\left[T_{_{\rm O}}^4({\bf r}_i)-T_{{\rm eq},\,\!i}^4\right]}{\bar{\Sigma}_i^2\,\bar{\kappa}_{_{\rm R}}(\rho_i,T_{{\rm eq},\,\!i})\,+\,\kappa_{_{\rm P}}^{-1}(\rho_i,T_{{\rm eq},\,\!i})}\;=\;0\,;
\end{eqnarray}
(b) the equilibrium internal energy, $u_{{\rm eq},\,\!i}=u(\rho_i,T_{{\rm eq},\,\!i})$; and (c) the thermalization timescale,
\begin{eqnarray}\label{EQN:TIMESCALE}
t_{{\rm therm},\,\!i}&=&\left\{u_{{\rm eq},\,\!i}-u_i\right\}\left\{\left.\frac{du_i}{dt}\right|_{_{\rm HYDRO}}+\left.\frac{du_i}{dt}\right|_{_{\rm RAD}}\right\}^{-1}\,.
\end{eqnarray}
We then advance $u_i$ through a time step $\Delta t$ by putting
\begin{eqnarray}\label{EQN:ADVANCE}
u_i(t\!+\!\Delta t)&=&u_i(t)\,\exp\left[\frac{-\,\Delta t}{t_{{\rm therm},\,\!i}}\right]\,+\,u_{{\rm eq},\,\!i}\left\{1-\exp\left[\frac{-\,\Delta t}{t_{{\rm therm},\,\!i}}\right]\right\}\!.
\end{eqnarray}

If $\Delta t\ll t_{{\rm therm},\,\!i}$, we are in a situation where thermal inertia effects are important, and Eqn. (\ref{EQN:ADVANCE}) approximates to
\begin{eqnarray}
u_i(t+\Delta t)&\simeq&u_i(t)\,+\,\left[u_{{\rm eq},\,\!i}\,-\,u_i(t)\right]\,\frac{\Delta t}{t_{{\rm therm},\,\!i}}\,;
\end{eqnarray}
i.e. in one dynamical timestep, $\Delta t$, the gas can only relax a little  towards thermal equilibrium.

On the other hand, if $\Delta t \gg t_{{\rm therm},\,\!i}$, Eqn. (\ref{EQN:ADVANCE}) approximates to
\begin{eqnarray}
u_i(t+\Delta t)&\simeq&u_{{\rm eq},\,\!i}\,;
\end{eqnarray}
i.e. thermal processes are occurring much faster than dynamical ones, and the gas is always close to thermal equilibrium. Using the above procedure we capture thermal inertia effects, whilst  avoiding the use of very small timesteps.

\subsection{Method implementation}
%%%%%%%%%%%%%%%%%%%%%%%%%%%%%%%%%%

The method is easy to implement. In practice we do the following, for each SPH particle $\,i$, at each timestep:

\begin{enumerate}
\item Calculate the pseudo-mean column-density ${\bar{\Sigma}}_i$ from the density, $\rho_i$ and gravitational potential, $\psi_i$, using Eqn. (\ref{EQN:SIGMABAR}). (In a simulation which includes stars, we must neglect their contribution to $\psi_i$.)

\item Calculate the pseudo-mean opacity, $\bar{\kappa}_{_{\rm R}}(\rho_i,T_i)$ and the Planck-mean opacity, $\kappa_{_{\rm P}}(\rho_i,T_i)$, by interpolation on a look-up table\footnote{Tabulated pseudo-mean opacities and internal energies (see next Section) can be obtained by contacting D.Stamatellos@astro.cf.ac.uk}.

\item Calculate the compressive plus viscous heating rate, $\left.du_i/dt\right|_{_{\rm HYDRO}}$, using Eqn.~(\ref{EQN:COMPRESS}), and the radiative heating rate, $\left.du_i/dt\right|_{_{\rm RAD}}$, using Eqn.~(\ref{EQN:RADHEAT}).

\item Calculate the equilibrium temperature, $T_{{\rm eq},\,\!i}$, using Eqn.~(\ref{EQN:EQUILIB}) and the thermalization timescale, $t_{{\rm therm},\,\!i}$, using Eqn.~(\ref{EQN:TIMESCALE}).

\item Update the internal energy, $u_i$, using Eqn.~(\ref{EQN:ADVANCE}); and hence also advance the temperature, $T_i$.
\end{enumerate}

\subsection{Limitations of the method}
%%%%%%%%%%%%%%%%%%%%%%%%%%%%%%%%%%%%%%

Although the method is very efficient, it evidently has limitations, in particular:

\begin{enumerate}

\item Because the diffusion approximation is applied here globally (to the whole pseudo-cloud), the method cannot capture in detail the local nature of radiative heating and cooling in the optically thick regime (i.e. the fact that in reality fluid elements exchange heat directly with other fluid elements, within a few photon mean-free-paths).

\item Because the pseudo-cloud {\it Ansatz} predicates a spherical polytropic cloud, the method works best for configurations which approximate to spherical symmetry. For example in simulations of disc fragmentation it handles the condensations better than the background disc. Notwithstanding this, even in an unperturbed disc the method is reasonably accurate, as shown by its performance of the Hubeny test (Stamatellos \& Whitworth, in preparation).

\end{enumerate}

%%%%%%%%%%%%%%%%%%%%%%%%%%%%%%%%%%%%%%%%%%%% SECTION
\section{Gas and dust properties}\label{sec:dustgas}
%%%%%%%%%%%%%%%%%%%%%%%%%%%%%%%%%%%%%%%%%%%%%%%%%%%%

\subsection{Gas-phase chemical abundances}
%%%%%%%%%%%%%%%%%%%%%%%%%%%%%%%%%%%%%%%%%%

Although metals make essential contributions to the opacity, they make very little contribution to the equation of state. Therefore, for the purpose of treating the gas-phase chemistry, we assume that the gas is 70\% hydrogen and 30\% helium by mass: $X=0.7$, $Y=0.3$, $Z=0\,$. At low temperatures, hydrogen is molecular, but as the temperature increases it becomes dissociated and then ionised. At low temperatures, helium is neutral atomic, but as the temperature increass it becomes ionised, first to He$^{+}$, and then to He$^{++}$. The relative abundances of these constituents depend on the density, $\rho_i$, and the temperature, $T_i$, and are calculated using Saha equations (e.g. Black \& Bodenheimer 1975), with the simplifying assumption that the dissociation of H$_2$ is complete before ionization of H$^{\rm o}$ begins; and similarly, that the ionization of He$^{\rm o}$ is complete before the ionization of He$^+$ begins.

%%%%%%%%%%%%%%%%%%%%%%%%%%%%%%%%%%%%%%%%%%%%%%%%%%%%%% FIGURE
\begin{figure}[!h]
\centerline{\includegraphics[height=8.8cm,angle=-90]{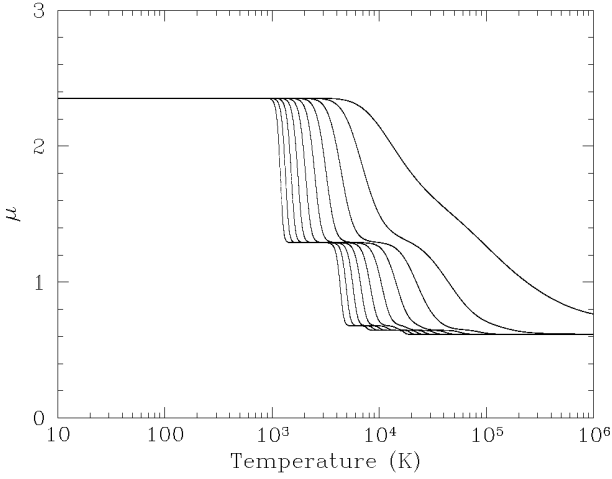}}
\caption{The variation of the mean molecular weight with density and temperature. Isopycnic curves are plotted from $\rho=10^{-18}\,{\rm g\, cm}^{-3}$ to $\rho=1 \,{\rm g\, cm}^{-3}$, every two orders of magnitude (bottom to top).}
\label{fig:mu}
\end{figure}
%%%%%%%%%%%%%%%%%%%%%%%%%%%%%%%%%%%%%%%%%%%%%%%%%%%%%%%%%%%%%

\subsection{The equation of state}
%%%%%%%%%%%%%%%%%%%%%%%%%%%%%%%%%%

If we define $y=n_{{\rm H}^{\rm o}}/2n_{{\rm H}_2}$ to be the degree of dissociation of hydrogen, $x=n_{{\rm H}^+}/n_{{\rm H}^{\rm o}}$ to be the degree of ionization of hydrogen, $z_1=n_{{\rm He}^+}/n_{{\rm He}^{\rm o}}$ to be the degree of single ionisation of helium, and $z_2=n_{{\rm He}^{++}}/n_{{\rm He}^+}$ to be the degree of double ionisation of helium, then the mean molecular weight is given by
\begin{eqnarray}
\mu_i&\!=\!&\mu(\rho_i,T_i)\,=\,\left[(1+y+2xy)\,\frac{X}{2}+(1+z_1+z_1z_2)\,\frac{Y}{4}\right]^{-1}\!.
\end{eqnarray}
Note that $(x,y,z_1,z_2)$ must be evaluated afresh for each SPH particle; the index $i$ has been dropped purely for simplicity. The variation of the mean molecular weight with density and temperature is shown in Fig.~\ref{fig:mu}.

For densities up to $\sim 0.03\,{\rm g\, cm}^{-3}$ the ideal gas approximation holds, and hence the gas pressure is
\begin{eqnarray}
P_i&=&\frac{\rho_i\,k_{_{\rm B}}\,T_i}{\mu_i\,m_{\rm H}}\,. 
\end{eqnarray}

%%%%%%%%%%%%%%%%%%%%%%%%%%%%%%%%%%%%%%%%%%%%%%%%%%%%%%% FIGURE
\begin{figure}
\centerline{\includegraphics[height=8.8cm,angle=-90]{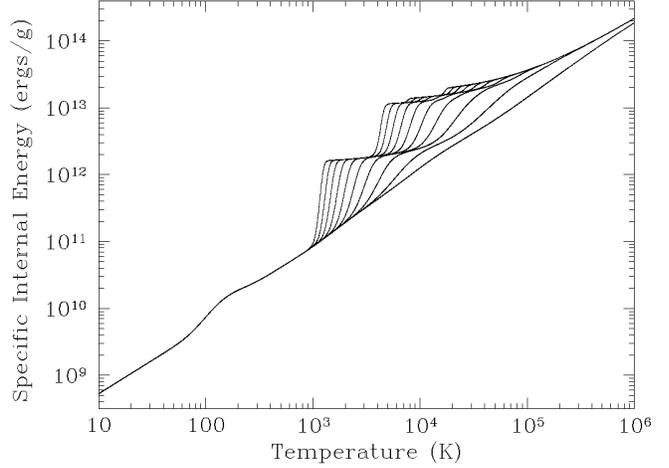}}
\caption{The variation of the specific internal energy with density and temperature. Isopycnic curves are plotted from $\rho=10^{-18}\,{\rm g\, cm}^{-3}\;{\rm to}\;\rho=1 \,{\rm g\, cm}^{-3}$, every two orders of magnitude (from top to bottom). The rotational degrees of freedom of H$_2$ are excited around $100\;{\rm to}\;200\,{\rm K}$, H$_2$ starts to be dissociated around $1,000\;{\rm to}\;10,000\,{\rm K}$, and H$^{\rm o}$ starts to be ionised around $4,000\;{\rm to}\;40,000\,{\rm K}$.}
\label{fig:ene}
\end{figure}
%%%%%%%%%%%%%%%%%%%%%%%%%%%%%%%%%%%%%%%%%%%%%%%%%%%%%%%%%%%%%%

\subsection{Specific internal energy of the gas}\label{sec:eos}
%%%%%%%%%%%%%%%%%%%%%%%%%%%%%%%%%%%%%%%%%%%%%%%%%%%%%%%%%%%%%%%

The specific internal energy (energy per unit mass) of an SPH particle $i$ is the sum of contributions from molecular, atomic and ionised hydrogen, atomic, singly ionised and doubly ionised helium, and the associated dissociation and ionisation energies,
\begin{equation}
u_i=u_{{\rm H}_2}\!+\!u_{\rm H}\!+\!u_{\rm He}\!+\!u_{\rm H_2\,\!DISS}\!+\!u_{\rm H\,ION}\!+\!u_{\rm He\,ION}\!+\!u_{\rm He^+\,\!ION},
\end{equation}
where
\begin{eqnarray}
u_{\rm H_2}&=&X(1-y)\,\left[\frac{3}{2}+c_i(T_i)\right]\,\frac{k_{_{\rm B}}T_i}{2m_{\rm H}}\,,\\
u_{\rm H}&=&X\,y\,(1+x)\,\frac{3k_{_{\rm B}}T_i}{2m_{\rm H}}\,,\\
u_{\rm He}&=&{Y\,(1+z_1+z_1 z_2)}\,\frac{3k_{_{\rm B}}T_i}{8m_{\rm H}}\,,\\
u_{\rm H_2\,\!DISS}&=&X\,y\,\frac{{\cal D}_{\rm H_2\,\!DISS}}{2\,m_{\rm H}}\,,\\
u_{\rm H\,ION}&=&X\,x\,y\,\frac{{\cal I}_{\rm H\,ION}}{m_{\rm H}}\,,\\
u_{\rm He\,ION}&=&Y\,z_1\, (1-z_2)\,\frac{{\cal I}_{\rm He\,ION}}{4\,m_{\rm H}}\,,\\
u_{\rm He^+\,\!ION}&=&Y\,z_1\,z_2\,\frac{{\cal I}_{\rm He^+\,\!ION}}{4\,m_{\rm H}}\,,
\end{eqnarray}
Here, ${\cal D}_{\rm H_2\,\!DISS}=4.5\,{\rm eV}$ is the dissociation energy of H$_2$; ${\cal I}_{\rm H\,ION}=13.6\,{\rm eV}$, ${\cal I}_{\rm He\,ION}=24.6\,{\rm eV}$ and ${\cal I}_{\rm He^+\,\!ION}=54.4\,{\rm eV}$ are the ionisation energies of H$^{\rm o}$, He$^{\rm o}$ and He$^+$, respectively; and the function 
\begin{equation}
c_i(T_i)=\left(\frac{T_{_{\rm ROT}}}{T_i}\right)^2 f(T_i)+\left(\frac{T_{_{\rm VIB}}}{T_i}\right)^2\frac{\exp(T_{_{\rm VIB}}/T_i)}{\left[ \exp(T_{_{\rm VIB}}/T_i)-1\right]^2}\,,
\end{equation}
with $T_{_{\rm ROT}}=85.4\,{\rm K}$ and $T_{_{\rm VIB}}=6100\,{\rm K}$ accounts for the rotational and vibrational degrees of freedom of H$_2$. The function $f(T_i)$ depends on the relative abundances of ortho- and para-H$_2$; we assume a fixed ortho-to-para ratio of 3:1. The variation of the specific internal energy with density and temperature is shown in Fig.~\ref{fig:ene}.

\subsection{Opacity}\label{sec:opacity}
%%%%%%%%%%%%%%%%%%%%%%%%%%%%%%%%%%%%%%%

\begin{figure}
\centerline{\includegraphics[height=8.9cm,angle=-90]{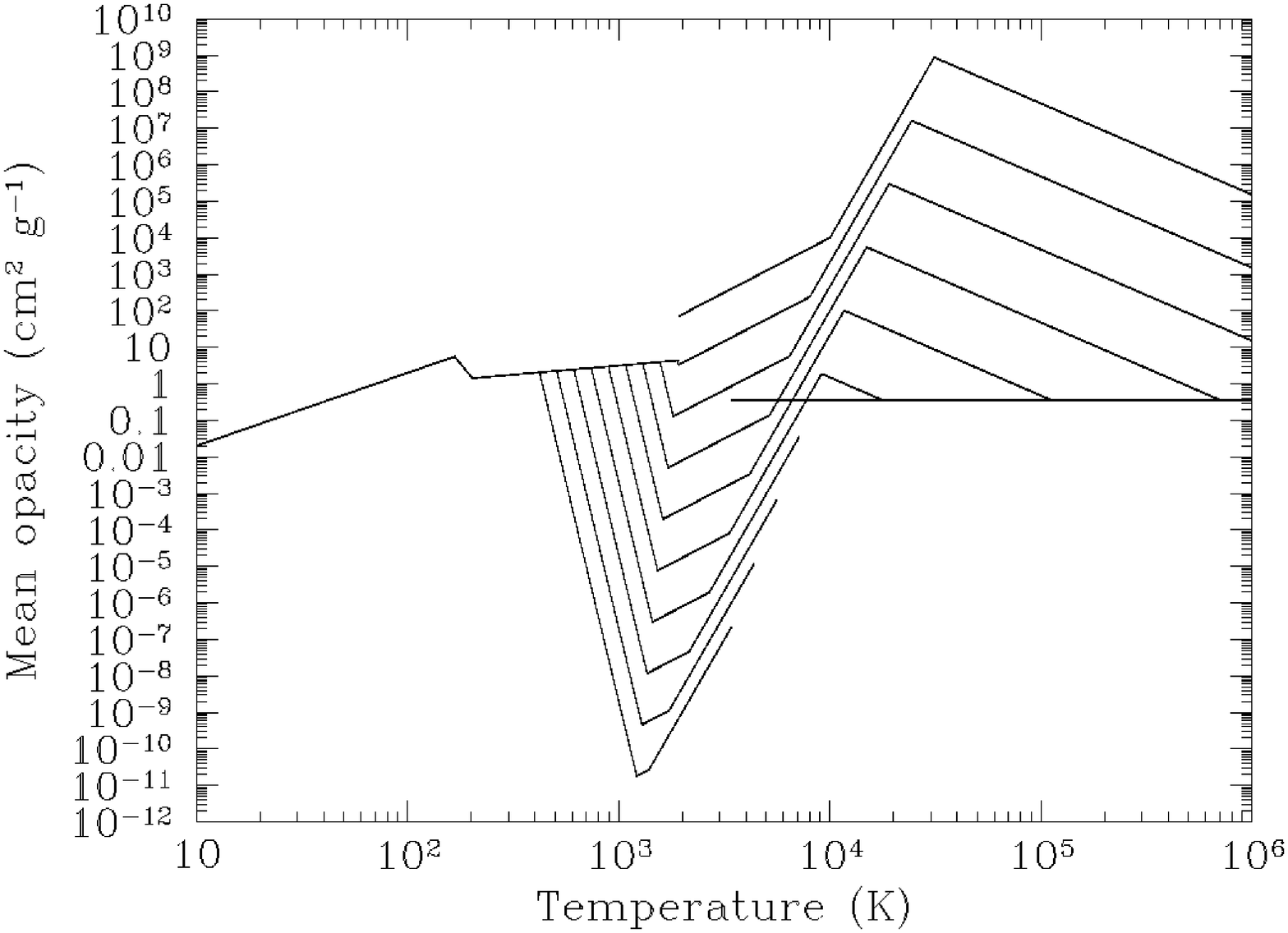}}
\caption{The variation of the {\it local Rosseland-mean opacity} with density and temperature.  Isopycnic curves are plotted from $\rho= 10^{-18}\,{\rm g\, cm}^{-3}$ to $\rho=1 \,{\rm g\, cm}^{-3}$, every two orders of magnitude (from bottom to top). The opacity gap is evident at temperatures $\sim 1,000\;{\rm to}\;3,000$~K, over a wide range of densities.}
\label{fig:opa}
\centerline{\includegraphics[height=8.9cm,angle=-90]{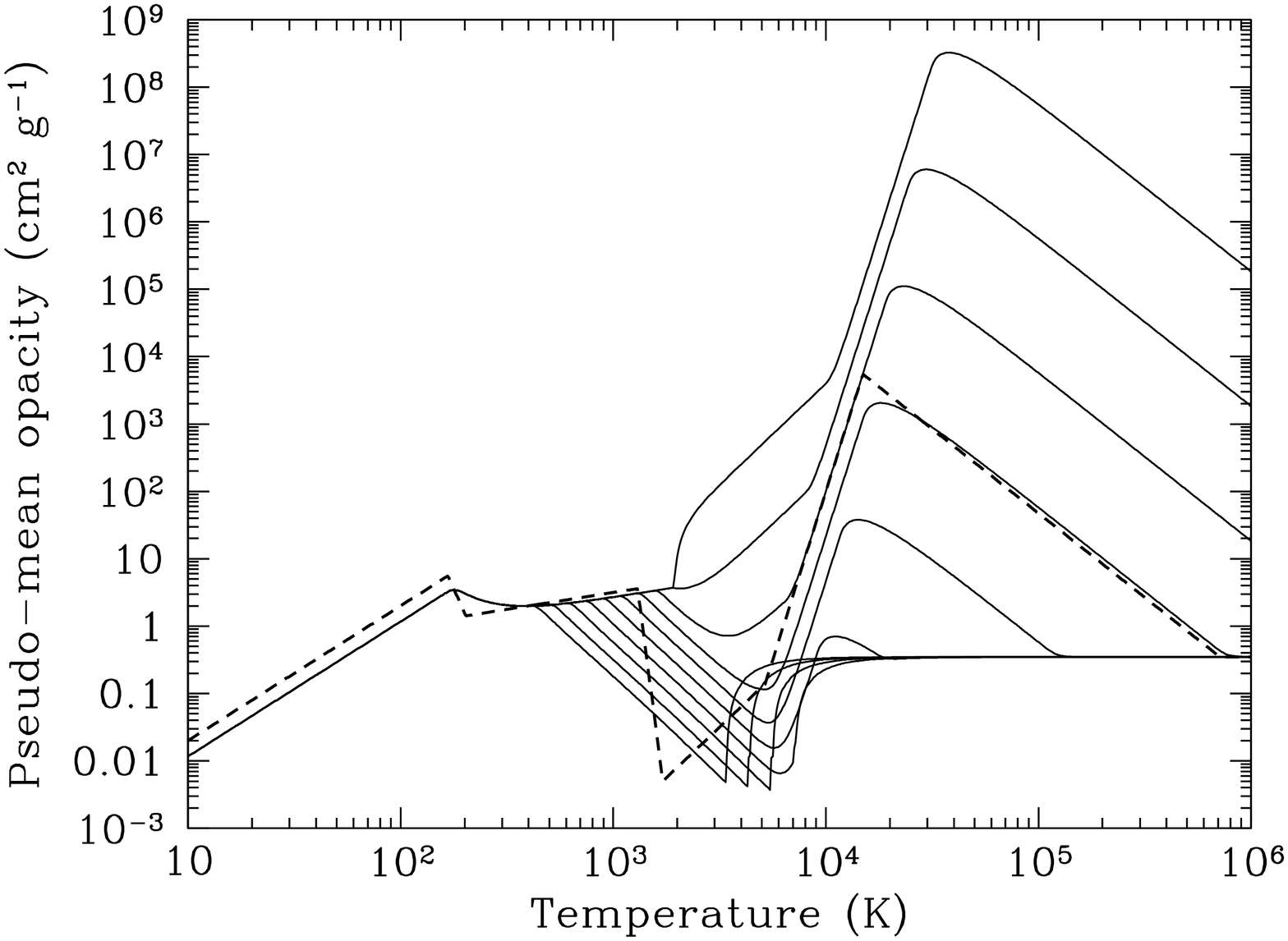}}
\caption{The variation with density and temperature of the {\it pseudo-mean opacity}. Isopycnic curves are plotted as in Fig.~\ref{fig:opa}. For comparison the {\it local} opacity  at density $\rho=10^{-6}\ {\rm g\ cm}^{-3}$ is also plotted (dashed line).}
\label{fig:nopa}
\end{figure}

In the present work we do not distinguish between the Rosseland-mean and Planck-mean opacities; we use the parametrisation proposed by Bell \& Lin (1994) for both, i.e.
\begin{equation} \label{eq:opacity}
\kappa_{_{\rm R}}(\rho,T)\;=\;\kappa_{_{\rm P}}(\rho,T)\;=\;\kappa_{_{\rm 0}}\,\rho^a\,T^b\,.
\end{equation}
Here $(\kappa_{_{\rm 0}},a,b)$ are constants which depend on the dominant physical process contributing to the opacity in different regimes of density and temperature (see Table~\ref{tab:opacity} and Fig.~\ref{fig:opa}).

The opacity at low temperatures is dominated by icy dust grains. At $T\sim 150$~K the ices evaporate and the opacity is due to metal grains up to $T\sim 1,000$~K, when the metal grains start to evaporate. The opacity drops considerably in the temperature range from $T\sim 1,000$~K to $T\sim 2,000$~K, as it is too hot for dust to exist, and too cool for H$^{-}$ to contribute, so the opacity is mainly due to molecules; this region of low opacity is sometimes referred to as the {\it opacity gap}. The opacity starts to increase again above $T\sim 2,000$~K due to H$^{-}$ absorption and then decreases again above $T\sim 10^4$~K, when free-free transitions take over. At very high temperatures, electron scattering delivers an approximately constant opacity.

At low temperatures, $T<2,000\,{\rm K}$, the Bell \& Lin parametrisation agrees well with the Rosseland-mean dust opacity calculated by Preibisch et al. (1993). Similarly, at high temperatures, $T>2,000\,{\rm K}$, it agrees well with the Rosseland-mean gas opacities calculated by Alexander \& Ferguson (1994) and Iglesias \& Rogers (1996).

Eqn.~(\ref{eq:opacity}) gives local Rosseland- and Planck-mean opacities. To calculate the pseudo-mean opacity used in Eqn. (\ref{EQN:RADHEAT}), we have to convolve this opacity with polytropic density and temperature profiles according to Eqn. (\ref{EQN:MEANOPACITY}). In Fig.~\ref{fig:nopa} we present the pseudo-mean opacity computed in this way, using a polytropic index $n=2$. We reiterate that the choice of $n$ affects the computed pseudo-mean opacities only weakly. 

%%%%%%%%%%%%%%%%%%%%%%%%%% TABLE
\begin{table}[!h]
\caption{Opacity law parameters (from Bell \& Lin 1994)} 
\centering
\renewcommand{\footnoterule}{}  % to avoid a line before footnotes
\begin{tabular}{clccc}\hline
\noalign{\smallskip}
 & Dominant opacity component  & $\kappa_{_{\rm 0}}$ & $a$ & $b$ \\
 & or physical process &&& \\
\noalign{\smallskip}\hline 
\noalign{\smallskip}
1 & Ice grains & $2\times10^{-4}$ & 0 & 2 \\
2 & Evaporation of ice grains & $2\times10^{16}$ & 0 & -7 \\
3 & Metal grains & 0.1 & 0 & 1/2 \\
4 & Evaporation of metal grains & $2\times10^{81}$ & 1 & -24 \\
5 & Molecules & $10^{-8}$ & 2/3 & 3 \\
6 & H$^{-}$ absorption & $10^{-36}$ & 1/3 & 10 \\ 
7 & bf and ff transitions  & $1.5\times10^{20}$ & 1 & -5/2 \\ 
8 & Electron scattering & 0.348 & 0 & 0 \\
\noalign{\smallskip}\hline
\end{tabular}
\label{tab:opacity}
\end{table}
%%%%%%%%%%%%%%%%%%%%%%%%%%%%%%%%

%%%%%%%%%%%%%%%%%%%%%%%%%%%%%%%%%%%%%%%%%%%%%%%%%%%%%%% FIGURE
%%%%%%%%%%%%%%%%%%%%%%%%%%%%%%%%%%%%%%%%%%%%%%%%%%%%%%%%%%%%%%

%%%%%%%%%%%%%%%%%%%%%%%%%%%%%%%%%%%%%%%%%%%%%%%%%%%%%%%%%%%%%%% SECTION
\section{The collapse of a 1-M$_{\sun}$ molecular cloud}\label{sec:mmi}
%%%%%%%%%%%%%%%%%%%%%%%%%%%%%%%%%%%%%%%%%%%%%%%%%%%%%%%%%%%%%%%%%%%%%%%

%%%%%%%%%%%%%%%%%%%%%%%%%%%%%%%%%%%%%%%%%%%%%%%%%%%%%%%%% FIGURE
\begin{figure*}
\centerline{\includegraphics[height=13.5cm,angle=-90]{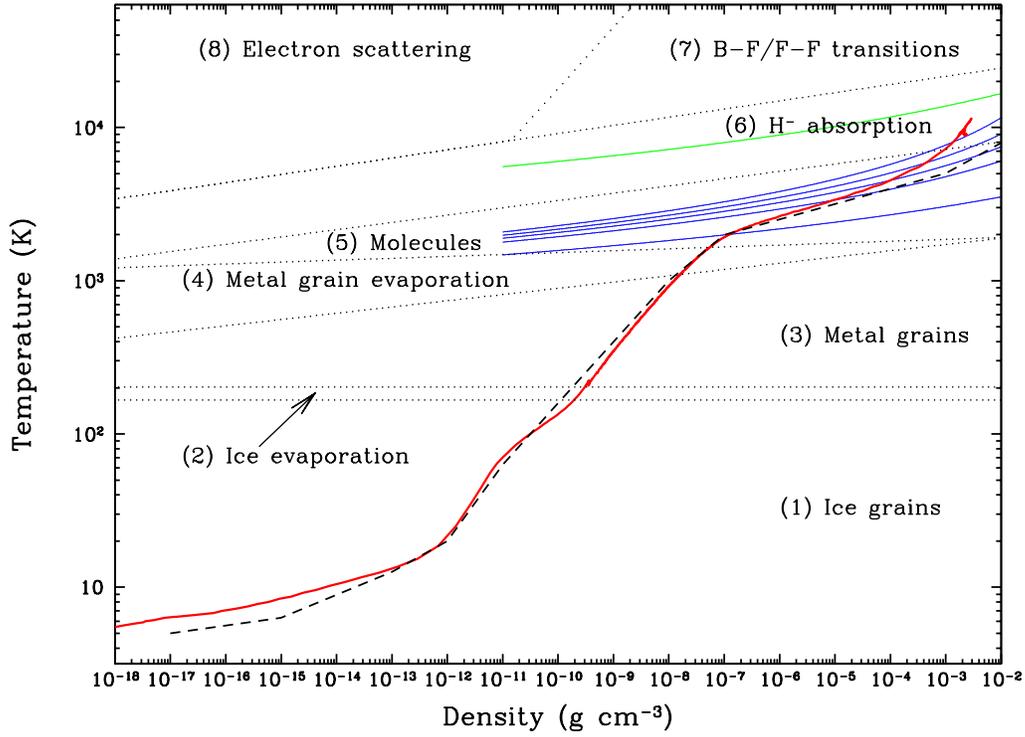}}
\caption{Evolution of the central density and central temperature of the collapsing cloud (thick red line). For comparison, the dashed black line shows the Masunaga \& Inutsuka (2000) simulation, and the dotted black lines delineate the different opacity regimes (see Fig.~\ref{fig:opa}). The thin solid lines are the loci for different degrees of H$_2$ dissociation (bottom set of blue lines, $y=0.01,\,0.2,\,0.4,\,0.6,\;{\rm and}\;0.8\,,$ respectively) and of H$^{\rm o}$ ionisation (top green line, $x=0.01$). The results of our model are very close to the simulation of Masunaga \& Inutsuka (2000). Differences at high densities are attributable to our using different opacities.}
\label{fig:td}
\end{figure*}
%%%%%%%%%%%%%%%%%%%%%%%%%%%%%%%%%%%%%%%%%%%%%%%%%%%%%%%%%%%%%%%%
%%%%%%%%%%%%%%%%%%%%%%%%%%%%%%%%%%%%%%%%%%%%%%%%%%%%%%%%%% FIGURE
\begin{figure*}
\centering\includegraphics[height=13.5cm,angle=-90]{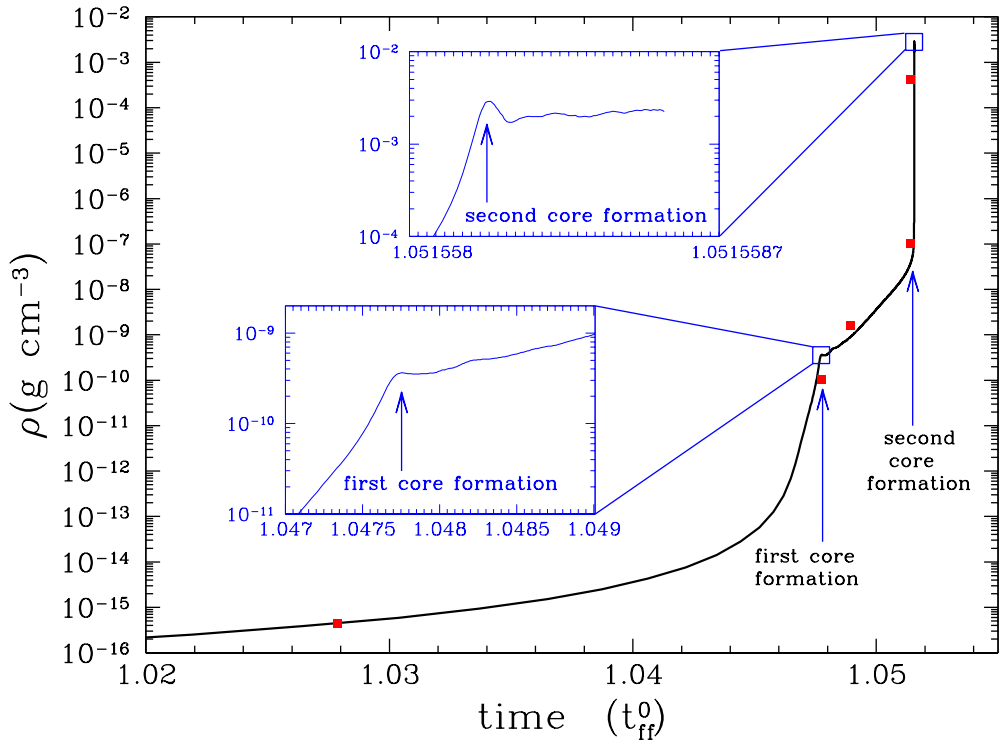}
\caption{Evolution of the central density with time, given in units of the initial free fall time $t_{\rm ff\,\!o}=1.781\times10^5\,{\rm yr}$. The times of the formation of the fist and second core are also marked. The red squares correspond to the Masunaga \& Inutsuka (2000) simulation.}
\label{fig:time0}
\end{figure*}
%%%%%%%%%%%%%%%%%%%%%%%%%%%%%%%%%%%%%%%%%%%%%%%%%%%%%%%%%%%%%%%%%

The first test of our new method for treating the energy equation in SPH is to simulate the collapse of a $1\,{\rm M}_\odot$ molecular cloud, which initially is spherical with radius $R=10^4\,{\rm AU}$ and uniform density $\rho_0=1.41\times10^{-19}\,{\rm g\, cm}^{-3}$. We set the background radiation temperature to $T_{_{\rm O}}({\bf r})=5\,{\rm K}$ This problem has been investigated by Masunaga \& Inutsuka (2000) using a code that treats the hydrodynamics in 1 dimension (i.e. assuming spherical symmetry) and the radiative transfer exactly in 3 dimensions (i.e.  by solving the angle-dependent and frequency dependent radiation transfer equation). It therefore constitutes a stiff test for our new method to reproduce their results. For the simulation presented here we use $2\times10^5$ SPH particles.

\subsection{Cloud collapse and the formation of the first and second cores}
%%%%%%%%%%%%%%%%%%%%%%%%%%%%%%%%%%%%%%%%%%%%%%%%%%%%%%%%%%%%%%%%%%%%%%%%%%%

The evolution of the cloud is followed up to density $\rho\sim10^{-3}\,{\rm g\, cm}^{-3}$ and temperature $T\sim 10^4$~K, i.e.  a difference from the initial conditions of about 16 orders of magnitude in density, and 3 orders of magnitude in temperature. 

As long as the density is below $\sim 10^{-12}\,{\rm g}\,{\rm cm}^{-3}$, the temperature increases slowly with increasing density. For $10^{-18}\,{\rm g}\,{\rm cm}^{-3}\la\rho\la10^{-13}\,{\rm g}\,{\rm cm}^{-3}$ we can approximate this with 
\begin{eqnarray}
T&\simeq&5\,{\rm K}\,\left[\frac{\rho}{10^{-18}\,{\rm g}\,{\rm cm}^{-3}}\right]^{0.08}
\end{eqnarray}
(cf. Larson 1973; Low \& Lynden-Bell 1976; Masunaga \& Inutsuka 2000; Larson 2005).

The cloud core starts heating more rapidly when it becomes optically thick. This continues until the the temperature reaches $T\sim 100\,{\rm K}$ at density $\rho\sim 3\times10^{-11}\,{\rm g\, cm}^{-3}$, when the rotational degrees of freedom of H$_2$ start to get excited, and hence the temperature increases more slowly with density (see the kink around $T\sim 100\,{\rm K}$ in Fig.~\ref{fig:td}). As the temperature increases further, the thermal pressure starts to decelerate the contraction, and the first core is formed (Larson 1969; Masunaga et al. 1998; Masunaga \& Inutsuka 2000; Whitehouse \& Bate 2006) at $t=1.048\,t_{\rm ff\,\!o}$, where $t_{\rm ff\,\!o}$ is the free-fall time at the start of collapse, i.e. $t_{\rm ff\,\!o}=\left[3\pi/(32\ G\ \rho_0)\right]^{1/2}=1.781\times10^5$~yr (see Fig.~\ref{fig:time0}).

The first core grows in mass, contracts, and heats up until the temperature reaches $T \sim 2,000\,{\rm K}$, when H$_2$ starts to dissociate. Consequently the compressional energy delivered by contraction does not all go to heat the core; instead some of it goes into dissociating H$_2$, and the second collapse starts. This second collapse proceeds until almost all of the molecular hydrogen at the centre has been dissociated. When the density reaches $\rho\sim 10^{-3}\,{\rm g\, cm}^{-3}$ and the temperature rises above $T\sim 10,000\,{\rm K}$, the collapse again decelerates and the second core (i.e. the protostar) is formed (Larson 1969; Masunaga \& Inutsuka 2000) at $t=1.052\ t_{\rm ff\,\!o}$. At first, the second core pulsates (see Fig~\ref{fig:time0} and Larson 1969), but eventually it settles down into quasistatic contraction. Due to computational constraints, the evolution is not followed further.

The evolution of the core in our simulation is very similar to that obtained by Masunaga \& Inutsuka (2000), as shown in Fig.\ref{fig:td}. Differences at densities $\stackrel{>}{_\sim}5\times10^{-6}\,{\rm g\, cm}^{-3}$ are attributable to the different opacities we use. The timescales in our simulation are also very similar to those obtained by Masunaga \& Inutsuka (2000), as is shown  in Fig.~\ref{fig:time0}, where the evolution of the density at the centre of the cloud  is plotted against time. The times  computed from our simulation fit well with the times from the  Masunaga \& Inutsuka (2000) simulation if we synchronise the two simulations at density $\rho=4.34\times10^{-13}\,{\rm g\, cm}^{-3}$, to avoid discrepancies due to small differences in the initial conditions at the onset of the collapse.

\subsection{Snapshots during the cloud collapse}
%%%%%%%%%%%%%%%%%%%%%%%%%%%%%%%%%%%%%%%%%%%%%%%%

In order to describe the evolution away from the centre of the cloud, and to make a more detailed comparison with the results of the Masunaga \& Inutsuka (2000) simulation, we focus on eight representative instants during the cloud evolution. Critical parameters at these instants are listed in Table~\ref{tab:elapsed}. The central densities have been chosen so as to match those used by Masunaga \& Inutsuka (2000) for the same purpose (see their Fig.~1 and their Table 1).

%%%%%%%%%% TABLE
\begin{table}
\caption{Colour code, elapsed time ($t$, measured from the beginning of the simulation and in units of the initial freefall time), central density ($\rho_{_{\rm C}}$), and central temperature ($T_{_{\rm C}}$), for the instants illustrated in Figs. \ref{fig:mmi.td} to \ref{fig:mmi.vcomp}.}
\label{tab:elapsed}\centering
\renewcommand{\footnoterule}{}  % to avoid a line before footnotes
\begin{tabular}{ccccr}\hline\noalign{\smallskip}
Label  & Colour & $t/t_{\rm ff\,\!o}$ & $\rho/{\rm g}\,{\rm cm}^{-3}$ & $T/{\rm K}$ \\
\noalign{\smallskip}\hline \noalign{\smallskip}
 1  & -      &$5.22\times 10^{-2}  $ 	&  $3.16\times 10^{-19}$&  5.0\\
 2  & black  &$8.43\times 10^{-1}  $ 	&  $1.68\times 10^{-18}$&  5.7\\
 3  & green  &$1.03038936$ 		&  $5.93\times 10^{-16}$&  8.0\\
 4  & cyan   &$1.04755301$ 		&  $1.09\times 10^{-10}$&  140\\
 5  & magenta&$1.04940519$ 		&  $1.59\times 10^{-9}$&   440\\
 6  & black  &$1.05153974$ 		&  $1.00\times 10^{-7}$&  1270\\
 7  & green  &$1.05155811$ 		&  $3.30\times 10^{-4}$&  5530\\
 8  & red    &$1.05155816$ 		&  $2.31\times 10^{-3}$& 10210\\
\noalign{\smallskip}\hline
\end{tabular}\end{table}
%%%%%%%%%%%%%%%%

%%%%%%%%%%%%%%%%%%%%%%%%%%%%%%%%%%%%%%%%%%%%%%%%%%%%%%%%%%% FIGURE
\begin{figure}[b]
\centerline{\includegraphics[height=8.6cm,angle=-90]{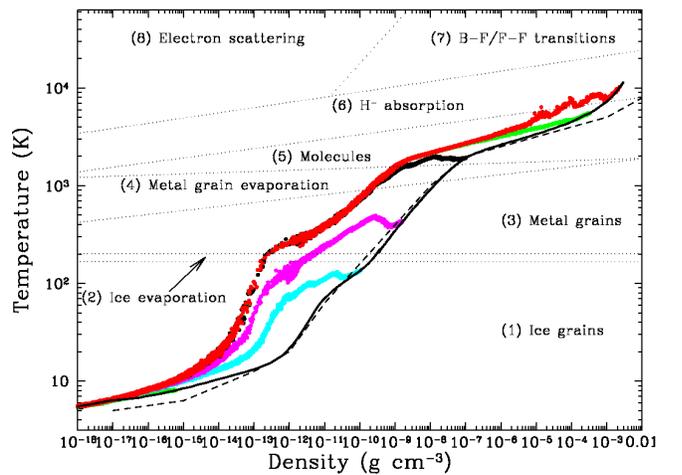}}
\caption{The run of temperature against density at different instants during the cloud evolution (instants 2 to 8 in Table~\ref{tab:elapsed}). The region around the centre of the cloud heats at lower densities than the centre of the cloud. The density increases as time evolves (black, green, cyan, magenta, black, green, red). For reference, we also plot the evolution of the temperature and density at the centre of the cloud in our simulation (lower solid black line) and in the Masunaga \& Inutsuka (2000) simulation (dashed black line).}
\label{fig:mmi.td}
\end{figure}
%%%%%%%%%%%%%%%%%%%%%%%%%%%%%%%%%%%%%%%%%%%%%%%%%%%%%%%%%%%%%%%%%%

%%%%%%%%%%%%%%%%%%%%%%%%%%%%%%%%%%%%%%%%%%%%%%%%%%%%%%%%%%%%%% FIGURE
\begin{figure}
\centerline{\includegraphics[height=8.6cm,angle=-90]{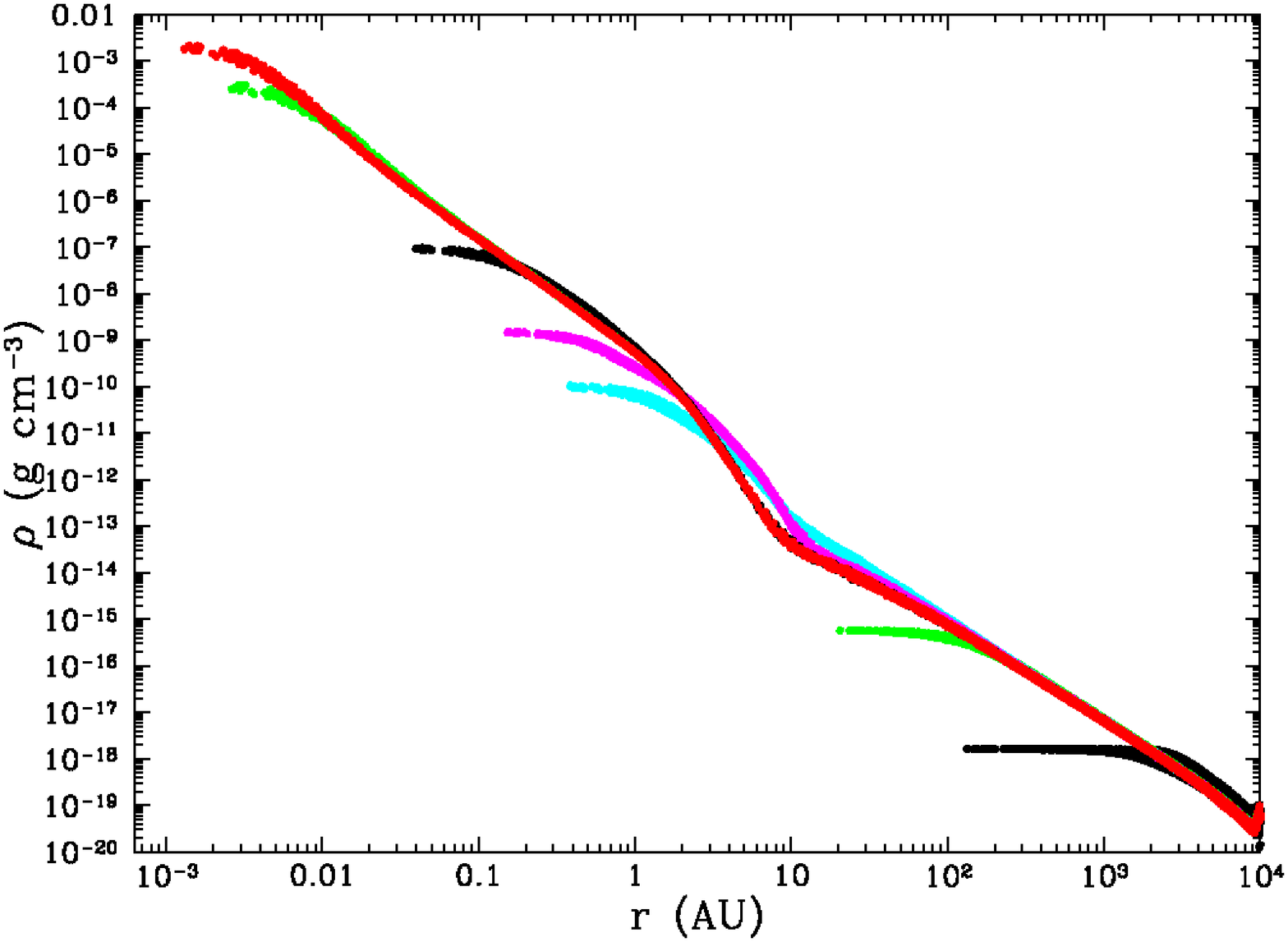}}
\caption{Density profiles at different instants during the cloud evolution (instants 2 to 8 in Table~\ref{tab:elapsed}). The colour coding is the same as in Fig.~\ref{fig:mmi.td}.}
\label{fig:mmi.dcomp}
\centerline{\includegraphics[height=8.6cm,angle=-90]{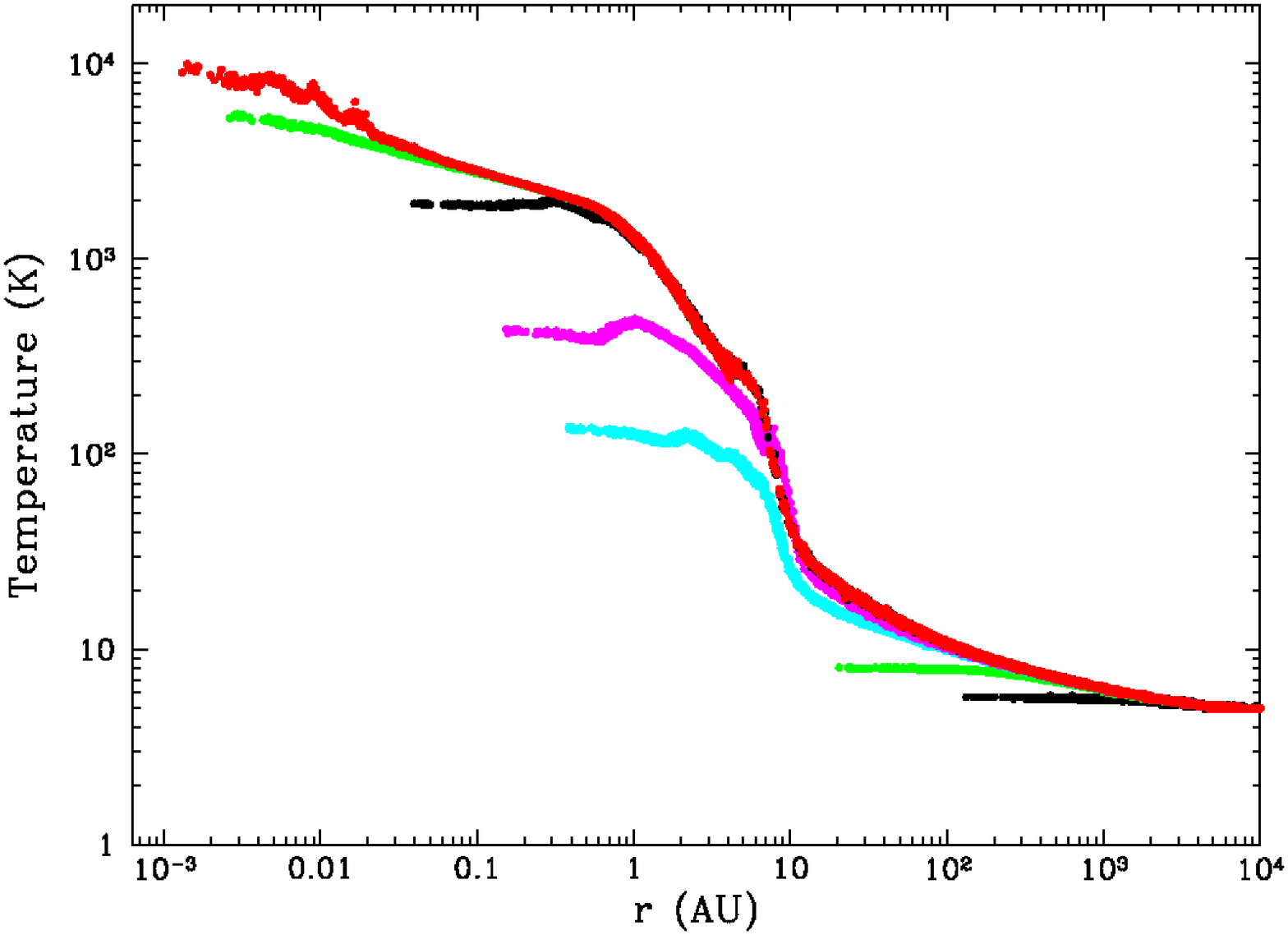}}
\caption{Temperature profiles at different instants during the cloud evolution (instants 2 to 8 in Table~\ref{tab:elapsed}). The colour coding is the same as in Fig.~\ref{fig:mmi.td}.}
\label{fig:mmi.tcomp}
\centerline{\includegraphics[height=8.7cm,angle=-90]{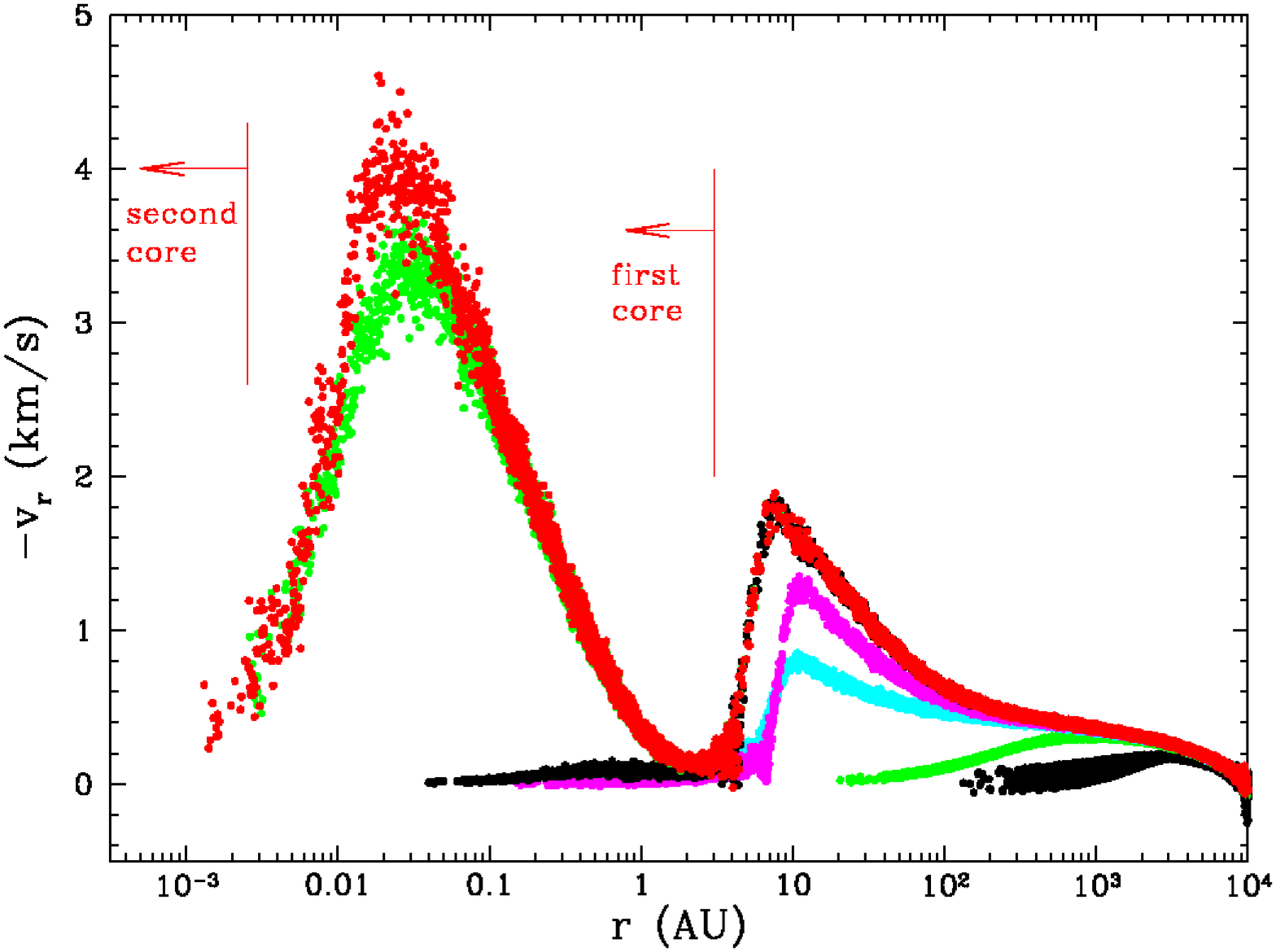}}
\caption{Radial infall velocity profiles at different instants during the cloud evolution (instants 2 to 8 in Table~\ref{tab:elapsed}). The colour coding is the same as in Fig.~\ref{fig:mmi.td}.}
\label{fig:mmi.vcomp}
\end{figure}
%%%%%%%%%%%%%%%%%%%%%%%%%%%%%%%%%%%%%%%%%%%%%%%%%%%%%%%%%%%%%%%%%%%%%

Fig.~\ref{fig:mmi.td} shows the run of temperature against density at the instants defined in Table~\ref{tab:elapsed}. During the early stages of the collapse the variation of temperature with density mimics the evolution of the central temperature and density (see the green points in Fig.~\ref{fig:mmi.td}; points representing previous instants are overlapped). At later stages, the regions around the centre start heating at lower densities, as in the  simulation of Whitehouse \& Bate (2006).

In Figs.~\ref{fig:mmi.dcomp} through \ref{fig:mmi.vcomp} we present the density, temperature and radial infall velocity profiles at different instants during the evolution of the cloud. These profiles are very similar to those reported by Masunaga \& Inutsuka (2000). The velocity profiles (Fig. \ref{fig:mmi.vcomp}) clearly show the formation of an accretion shock at the boundary of the first core at radius $R\sim 3\;{\rm to}\;5\,{\rm AU}$. There is also an accretion shock at the boundary of the second core, initially at a radius of $R\sim 0.003\,{\rm AU}$, but later expanding to $R\sim 0.01\,{\rm AU}$ (cf. Larson 1969).

\subsection{Convergence}
%%%%%%%%%%%%%%%%%%%%%%%%

\begin{figure}[!h]
\centerline{\includegraphics[height=8.8cm,angle=-90]{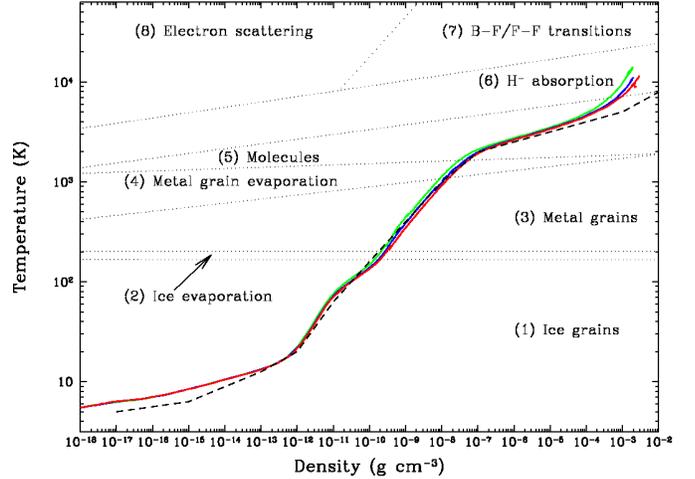}}
\caption{The run of central temperature against central density during the evolution of a $1\,{\rm M}_\odot$ molecular cloud, simulated using different numbers of SPH particles: ${\cal N}\simeq 2\times10^4\;{\rm (green)},\;5\times10^4\;{\rm (blue)},\;10^5\;{\rm (cyan)},\;{\rm and}\;2\times10^5\;{\rm (red)}$. The results converge for ${\cal N}\stackrel{>}{_\sim} 10^5$. The black dashed line represents the Masunaga \& Inutsuka (2000) simulation.}
\label{fig:td.conv}
\end{figure}

We repeat this simulation using different numbers of SPH particles, ${\cal N}$, to check for convergence. We use ${\cal N}\simeq 2\times10^4,\;5\times10^4,\;10^5,\;{\rm and}\;2\times10^5$ (Fig.~\ref{fig:td.conv}). The run of central temperature against central density is almost identical for different numbers of particles, and the results are fully converged up to densities $\rho\sim 0.003\,{\rm g\, cm}^{-3}$ with ${\cal N}\ga 10^5$. Currently available supercomputing facilities allow SPH simulations of star formation with up to $3\times 10^7$ particles, and so the convergence condition quoted above can easily be met.

%%%%%%%%%%%%%%%%%%%%%%%%%%%%%%%%%%%%%%%%%%%%%%%%%%%%%%%%%%% FIGURE
%%%%%%%%%%%%%%%%%%%%%%%%%%%%%%%%%%%%%%%%%%%%%%%%%%%%%%%%%%%%%%%%%%

%%%%%%%%%%%%%%%%%%%%%%%%%%%%%%%%%%% SECTION
\section{Additional tests}\label{sec:tests}
%%%%%%%%%%%%%%%%%%%%%%%%%%%%%%%%%%%%%%%%%%%

\subsection{Boss \& Myhill (1992)}\label{sec:bmtest}
%%%%%%%%%%%%%%%%%%%%%%%%%%%%%%%%%%%%%%%%%%%%%%%%%%%%

\begin{figure}
\centering{\includegraphics[height=8.9cm,angle=-90]{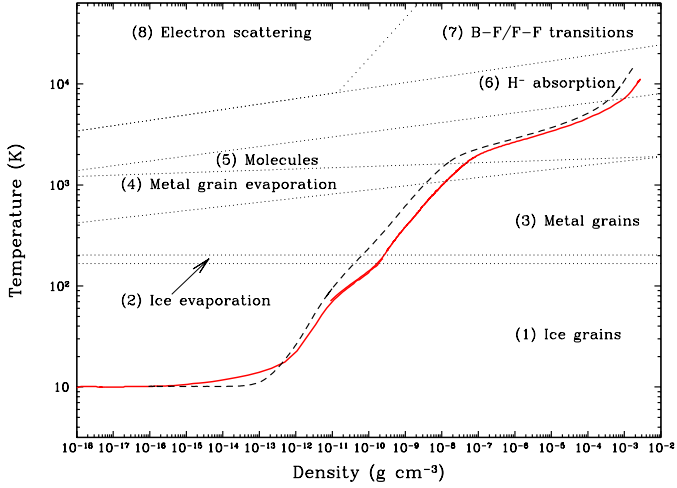}}
\caption{Evolution of the central density and central temperature for the Boss \& Myhill (1992) test problem. The dashed line corresponds to the Whitehouse \& Bate (2006) simulation, and the dotted lines define the different opacity regimes (see Fig.~\ref{fig:opa}). The results of our model are very close to the simulation of Whitehouse \& Bate (2006). Differences at densities $\stackrel{>}{_\sim}5\times10^{-6}\,{\rm g\, cm}^{-3}$ are attributable to the use of different opacities.}
\label{fig:td.bm}
\end{figure}

The second test of our new method is to simulate the evolution of a $1.2\,{\rm M}_\odot$ cloud with uniform initial density $\rho=1.7\times10^{-19}\,{\rm g\, cm}^{-3}$, uniform initial temperature $T=10\,{\rm K}$, and initial radius $R=1.5\times10^{17}\,{\rm cm}$, as originally investigated by Boss \& Myhill (1992). This problem has recently been revisited by Whitehouse \& Bate (2006), using SPH with flux-limited diffusion. In our simulation we have ${\cal N}\simeq 1.5\times10^5$ SPH particles. The evolution of the cloud is very similar to the evolution already described in Section \ref{sec:mmi}. Fig.~\ref{fig:td.bm} compares the run of central temperature against central density which we obtain, with that obtained by Whitehouse \& Bate (2006). There are three small differences. (i) In our simulation, the cloud starts heating before it becomes optically thick (as reported also by Masunaga \& Inutsuka (2000) in a similar test). In contrast, the Whitehouse \& Bate (2006) cloud remains strictly isothermal during this phase. (ii) In the Whitehouse \& Bate (2006) simulation the centre of the cloud heats up more rapidly at densities $\rho \sim 10^{-11}\;{\rm to}\;10^{-6}\,{\rm g\, cm}^{-3}$, i.e. at lower densities than in our simulation. However, Masunaga \& Inutsuka (2000) in a similar test also find lower temperatures than Whitehouse \& Bate (2006) in this regime. (iii) There are differences at densities $\rho \ga 5\times10^{-6}\,{\rm g\, cm}^{-3}$, which are attributable to the use of different opacities. These differences are small and the overall evolution of the cloud is similar in the two simulations.

%%%%%%%%%%%%%%%%%%%%%%%%%%%%%%%%%%%%%%%%%%%%%%%%%%%%%%%% FIGURE
%%%%%%%%%%%%%%%%%%%%%%%%%%%%%%%%%%%%%%%%%%%%%%%%%%%%%%%%%%%%%%%%

\subsection{Boss \& Bodenheimer (1979)}\label{sec:bbtest}
%%%%%%%%%%%%%%%%%%%%%%%%%%%%%%%%%%%%%%%%%%%%%%%%%%%%%%%%%

\begin{figure}
\centering{\includegraphics[height=8.9cm,angle=-90]{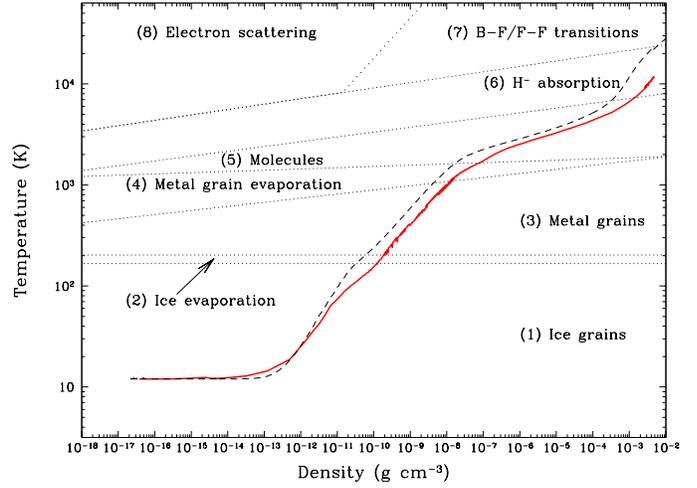}}
\caption{Evolution of the density and temperature at the densest part of a collapsing, rotating molecular cloud. The dashed line corresponds to the Whitehouse \& Bate (2006) simulation of the same problem. The results of our simulation are very close to those of Whitehouse \& Bate (2006). Differences at densities $\rho \stackrel{>}{_\sim}5\times10^{-6}\,{\rm g\, cm}^{-3}$ are attributable to different opacities. }
\label{fig:td.bb}
\end{figure}

The third test of our new method is to simulate the evolution of a rotating $1.2\,{\rm M}_\odot$ cloud; the cloud initially rotates as a solid body with angular velocity $\Omega=1.6\times10^{-12}\,{\rm rad}\,{\rm s}^{-1}$, and so the ratio of rotational-to-gravitational energy is $\beta=0.26$; the density includes an $m=2$ perturbation, i.e. $\rho=1.44\times 10^{-17}\,{\rm g\, cm}^{-3}[1+0.5\ \cos(2\phi)]$, where $\phi$ is the azimuthal angle in the plane perpendicular to the axis of rotation; the initial radius of the cloud is $R=3.2\times 10^{16}\,{\rm cm}$ and the temperature is initially uniform at $T=12\,{\rm K}$; this problem was originally investigated by Boss \& Bodenheimer (1979), and has recently been revisited by Whitehouse \& Bate (2006). In our simulation we have ${\cal N}\simeq 1.5\times10^5$ SPH particles. Fig.~\ref{fig:td.bb} compares the evolution of the density and temperature of the densest part of the cloud in our simulation, with that obtained by Whitehouse \& Bate (2006). The differences are again only small.

\begin{figure*}
\centering{\includegraphics[height=5.9cm,angle=-90]{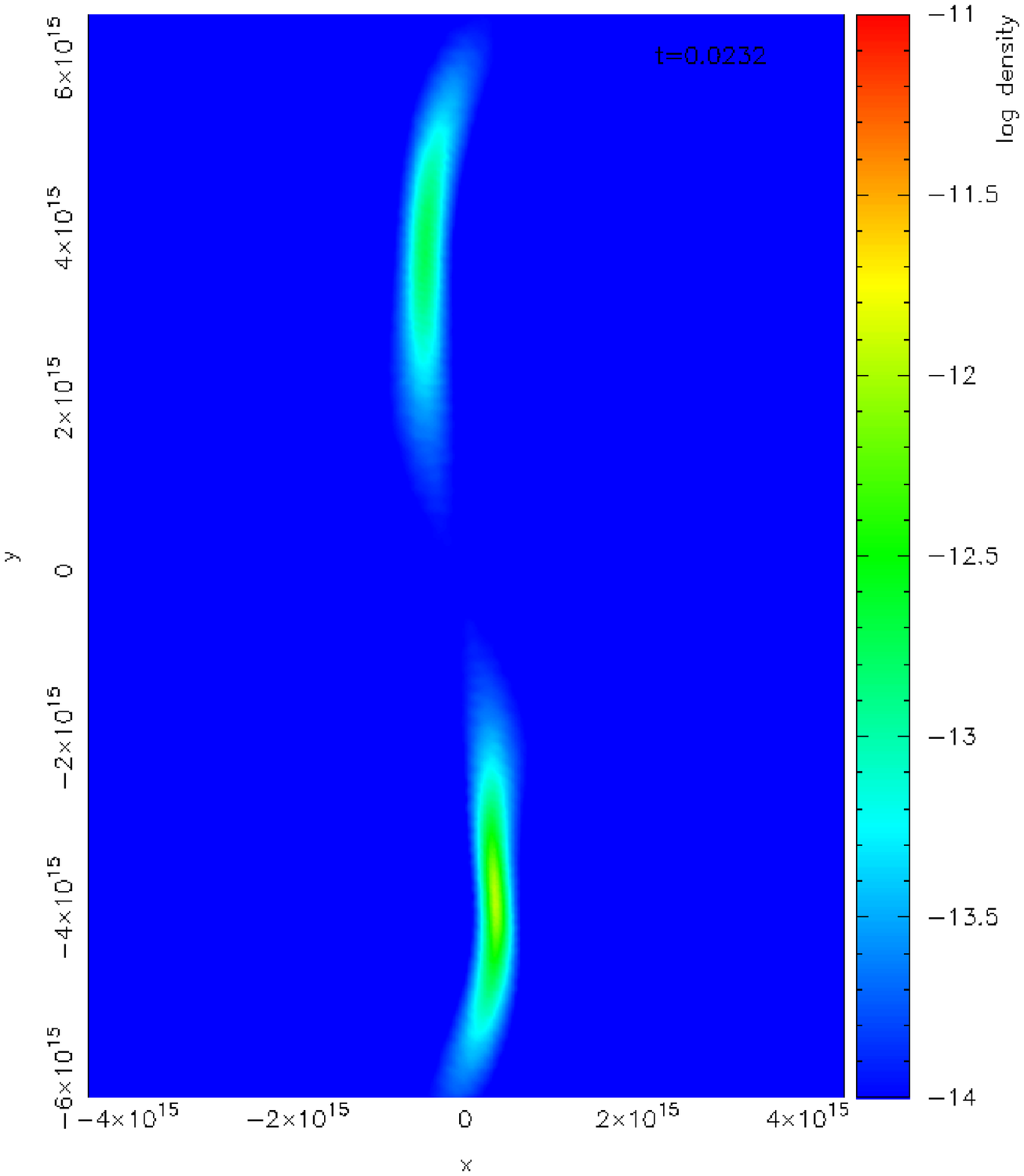}
\includegraphics[height=5.9cm,angle=-90]{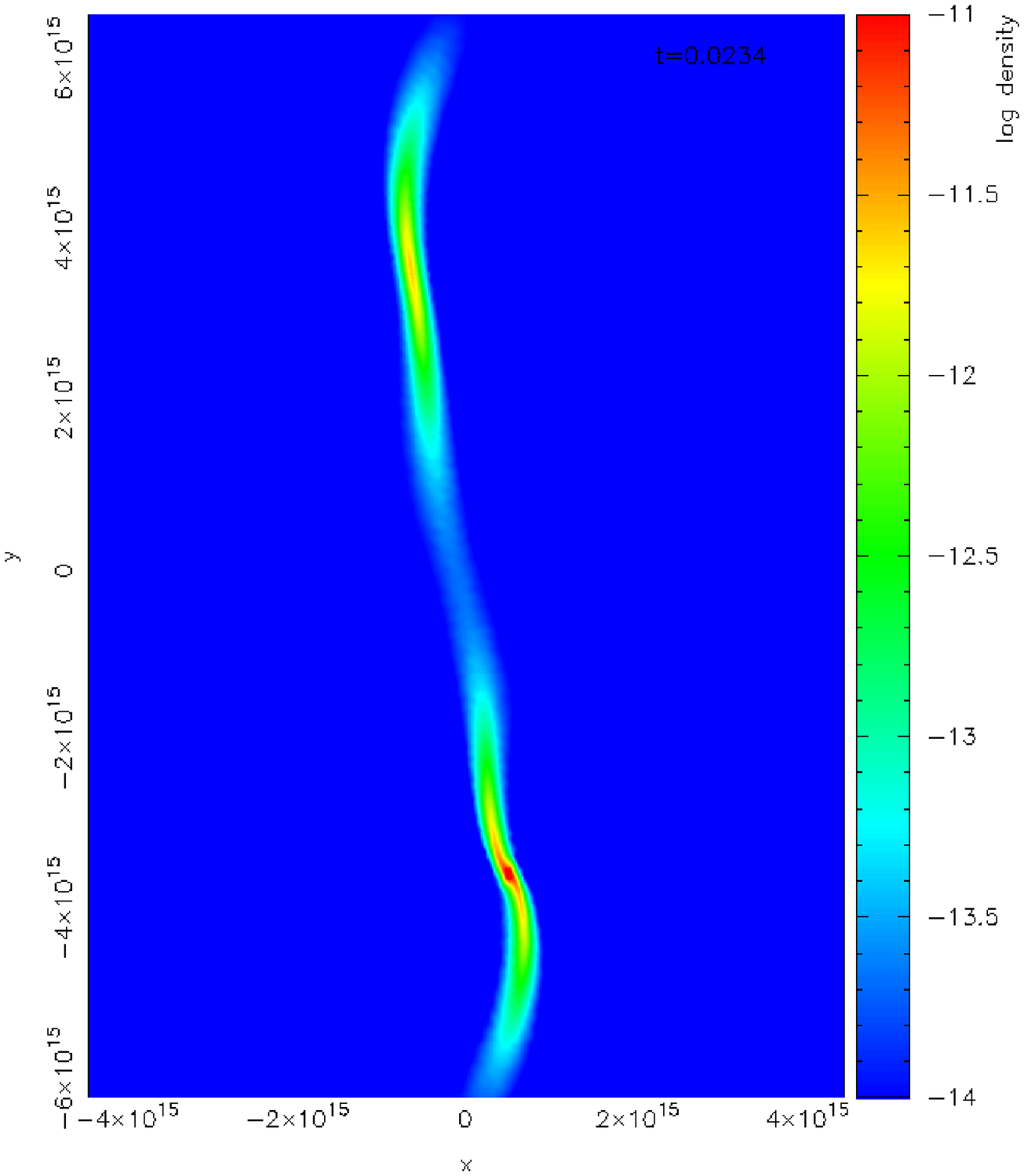}
\includegraphics[height=5.9cm,angle=-90]{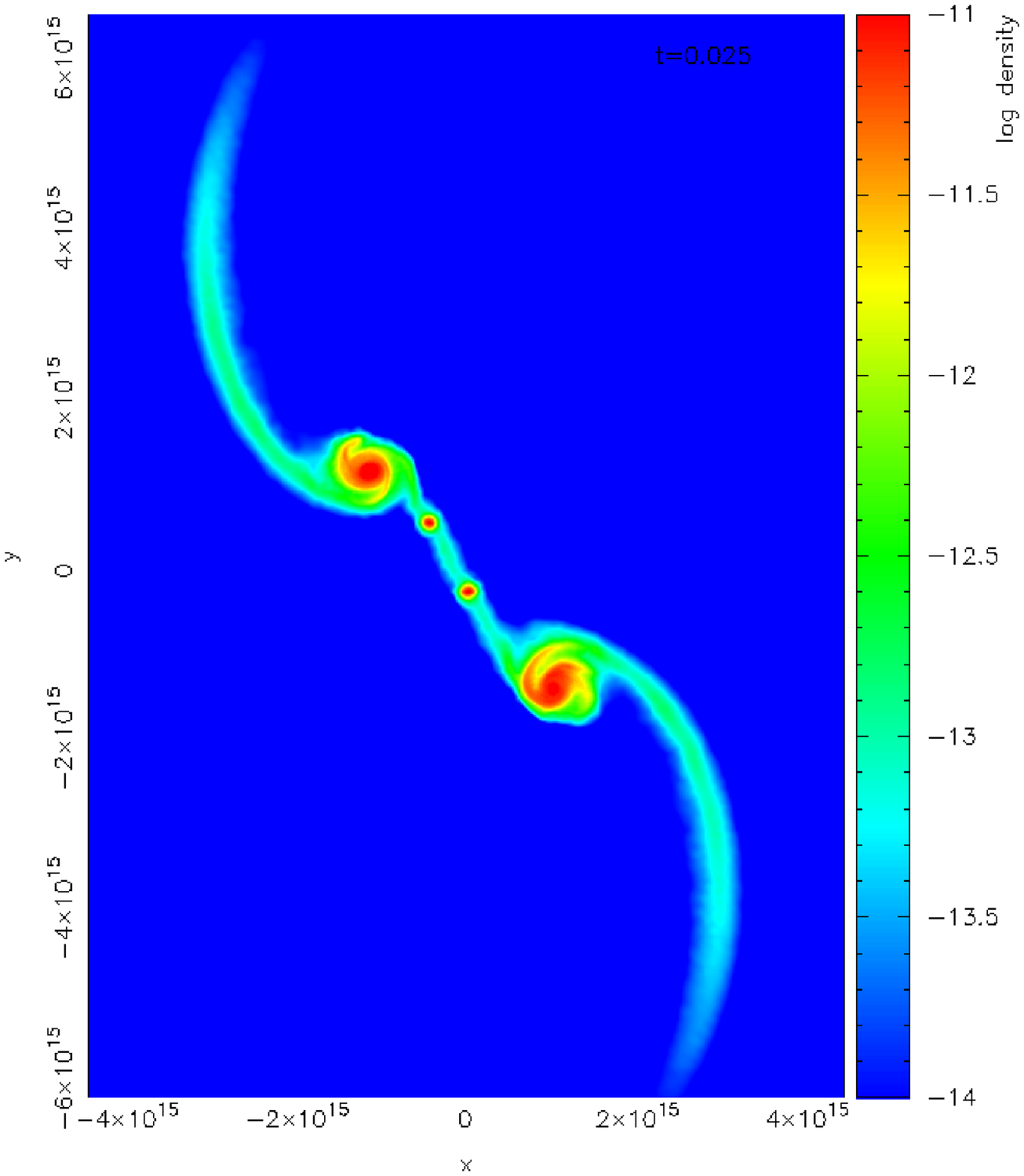}}
\centering{
\includegraphics[height=5.9cm,angle=-90]{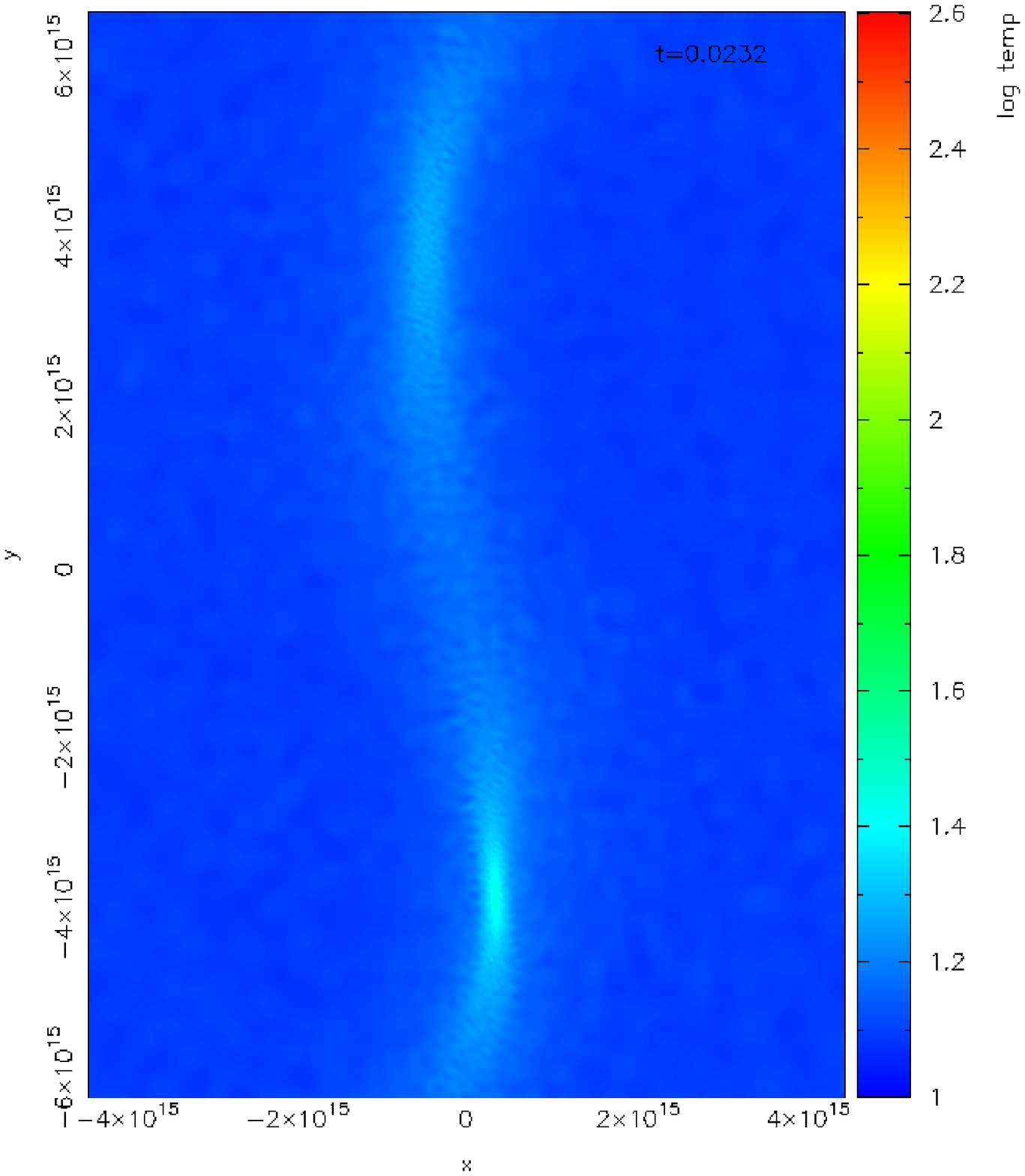}
\includegraphics[height=5.9cm,angle=-90]{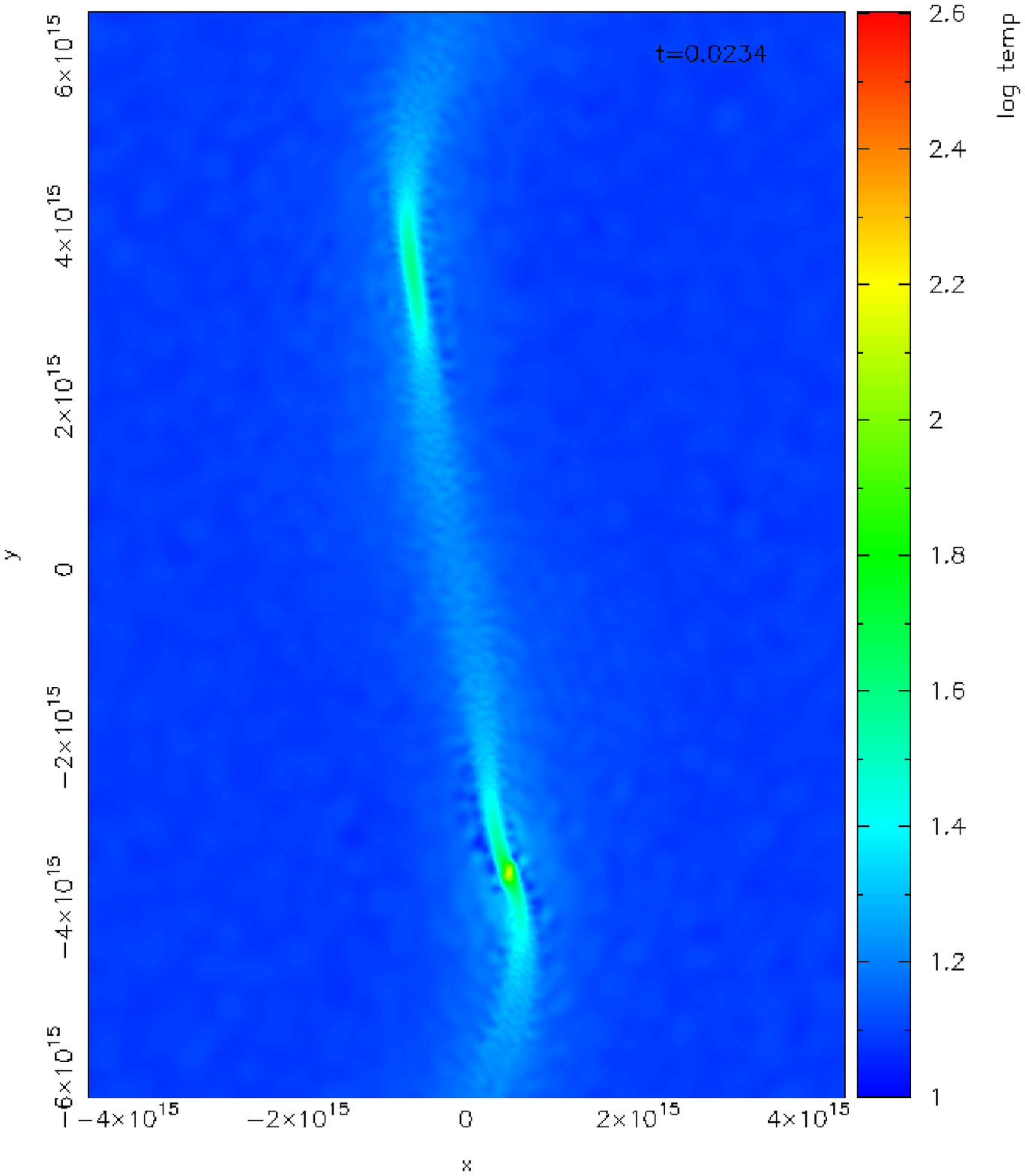}
\includegraphics[height=5.9cm,angle=-90]{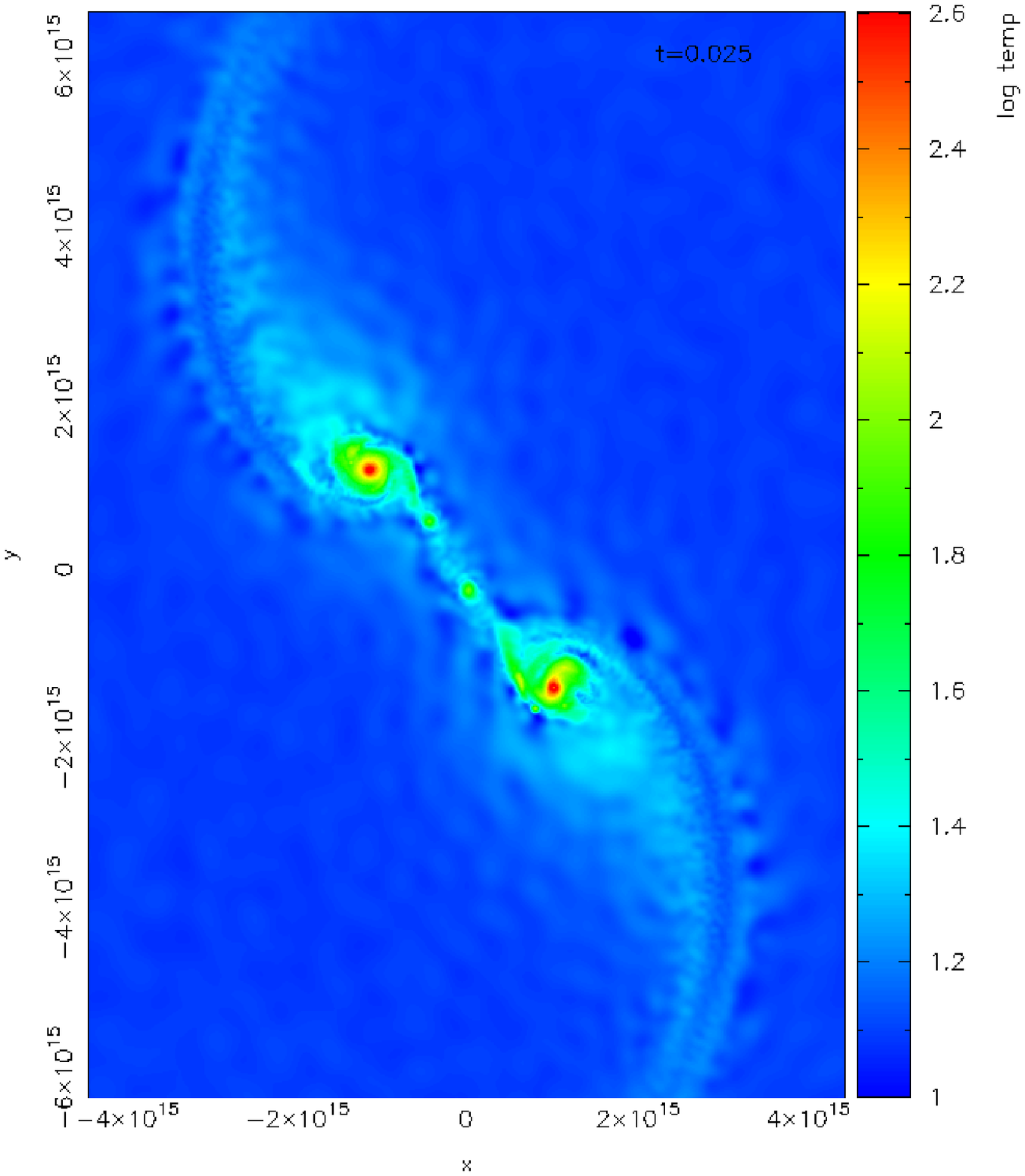}}
\caption{Three instants from our simulation of the Boss \& Bodenheimer test problem, at $t_1=0.02315\,{\rm Myr}$, $t_2=0.02343{\rm Myr}$, and $t_3=0.025\,{\rm Myr}$ (left to right). We plot the logarithmic density (top) and the logarithmic temperature (bottom), on the $xy$-plane, i.e. the plane perpendicular to the rotation axis. The central densities and central temperatures of the southern condensation are 
$\rho_{_{1{\rm S}}}=1.2\times10^{-12}\, {\rm g\, cm}^{-3}$, $T_{_{1{\rm S}}}=25\,$K, 
$\rho_{_{2{\rm S}}}=1.6\times10^{-10}\, {\rm g\, cm}^{-3}$, $T_{_{2{\rm S}}}=150\,$K,
$\rho_{_{3{\rm S}}}=3.8\times10^{-3}\, {\rm g\, cm}^{-3}$, $T_{_{3{\rm S}}}=10800\,$K.
The corresponding central densities and central temperatures of the northern condensation are
$\rho_{_{1{\rm N}}}=1.7\times10^{-13}\, {\rm g\, cm}^{-3}$, $T_{_{1{\rm N}}}=18\,$K,
$\rho_{_{2{\rm N}}}=2.4\times10^{-12}\, {\rm g\, cm}^{-3}$, $T_{_{2{\rm N}}}=35\,$K,
$\rho_{_{3{\rm N}}}=6.3\times10^{-10}\, {\rm g\, cm}^{-3}$, $T_{_{3{\rm N}}}=780\,$K.}
\label{fig:image.bb}
\end{figure*}

The result of the collapse is a binary with separation $S\sim 500\,{\rm AU}$. In Fig.~\ref{fig:image.bb} we plot the density and the temperature on the $xy$-plane at three instants during the evolution. The components of the binary system are connected by a bar which subsequently fragments. If the collapse were isothermal, this bar should not show any tendency to fragment (e.g. Truelove et al. 1998; Klein et al. 1999; Kitsionas \& Whitworth 2002). However, Bate \& Burkert (1997) show that if the gas is allowed to heat, but the heating happens at sufficiently high densities, $\rho\ga 0.3\times 10^{-13}\,{\rm g\, cm}^{-3}$, then the bar does fragment. In our simulation, the gas in the bar starts to heat up rapidly only when the density reaches $\rho\sim 7\times 10^{-13}\,{\rm g\, cm}^{-3}$. Therefore the fragmentation of the bar is consistent with the predictions of Bate \& Burkert (1997). However, we note that Whitehouse \& Bate (2006) do not report any bar fragmentation.

%%%%%%%%%%%%%%%%%%%%%%%%%%%%%%%%%%%%%%%%%%%%%%%%%%%%%%%%%%%%%%%
%%%%%%%%%%%%%%%%%%%%%%%%%%%%%%%%%%%%%%%%%%%%%%%%%%%%%%%%%%%%%%%

%%%%%%%%%%%%%%%%%%%%%%%%%%%%%%%%%%%%%%%%%%%%%%%%%%%%%%%%%%%%%%%%%
%%%%%%%%%%%%%%%%%%%%%%%%%%%%%%%%%%%%%%%%%%%%%%%%%%%%%%%%%%%%%%%%%

\subsection{Thermal relaxation}
%%%%%%%%%%%%%%%%%%%%%%%%%%%%%%%

Finally, we test the time-dependence of our new method by simulating the relaxation of temperature fluctuations in a static sphere with uniform density $\rho=10^{-19}{\rm g\ cm^{-3}}$ and radius $R=10,000\,{\rm AU}$. We assume an equilibrium temperature of $T_{_{\rm O}}=10~{\rm K}$ and an initial temperature perturbation of the form $\Delta T=\Delta T_{_{\rm O}}\sin(kr)/kr$ (Masunaga et al. 1998; Spiegel 1957), where $\Delta T_{_{\rm O}}=0.15\,{\rm K}$ is the amplitude of the perturbation and $k=\pi/(2500\,{\rm AU})$ is its characteristic wavenumber. Masunaga et al. (1998) have shown that at subsequent times the temperature should be
\begin{eqnarray}\label{EQN:RELAX}
T(r,t)&=&T_{_{\rm O}}\;+\;\Delta T_{_{\rm O}}\,\frac{\sin(kr)}{kr}\;{\rm e}^{-\omega(k)t}\,,
\end{eqnarray}
where 
\begin{eqnarray}\label{eq:rate}
\omega(k)&=&\gamma\,\left[1-\frac{\kappa_{_{\rm O}}}{k}\,\cot^{-1}\!\left(\frac{\kappa_{_{\rm O}}}{k} \right)\right]
\end{eqnarray}
is the relaxation rate, 
\begin{eqnarray}
\gamma&=&\frac{16\ \sigma_{\rm SB}\,\kappa_{_{\rm O}}\,T_{_{\rm 0}}^3}{\rho\,c_{_{\rm V}}}\,,
\end{eqnarray}
$\kappa_{_{\rm O}}$ is the opacity at the equilibrium temperature, and $c_{V}$ is the heat capacity of the material.

%%%%%%%%%%%%%%%%%%%%%%%%%%%%%%%%%%%%%%%%%%%%%%%%%%%%%%%% FIGURE
\begin{figure}[!h]
\centering{\includegraphics[height=8cm]{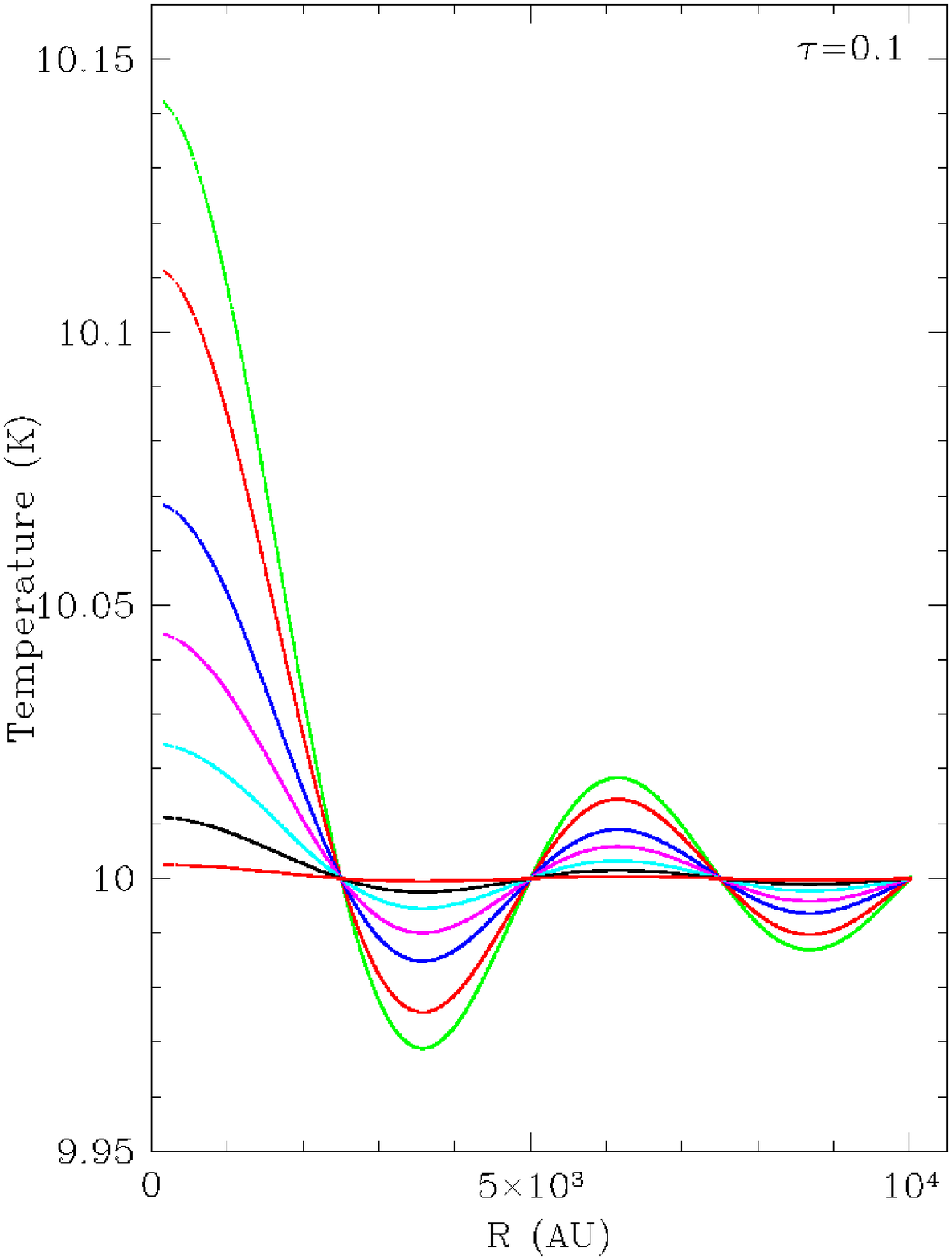}}
\caption{Thermal relaxation of a static, uniform sphere of optical depth $\tau=0.1$; seven instants are plotted, showing the temperature relaxing to its equilibrium value $T_{\rm eq}=10\,{\rm K}$.}
\label{fig:relax.tau=0.1}
%\end{figure}
%\begin{figure}
\centering{\includegraphics[height=8cm]{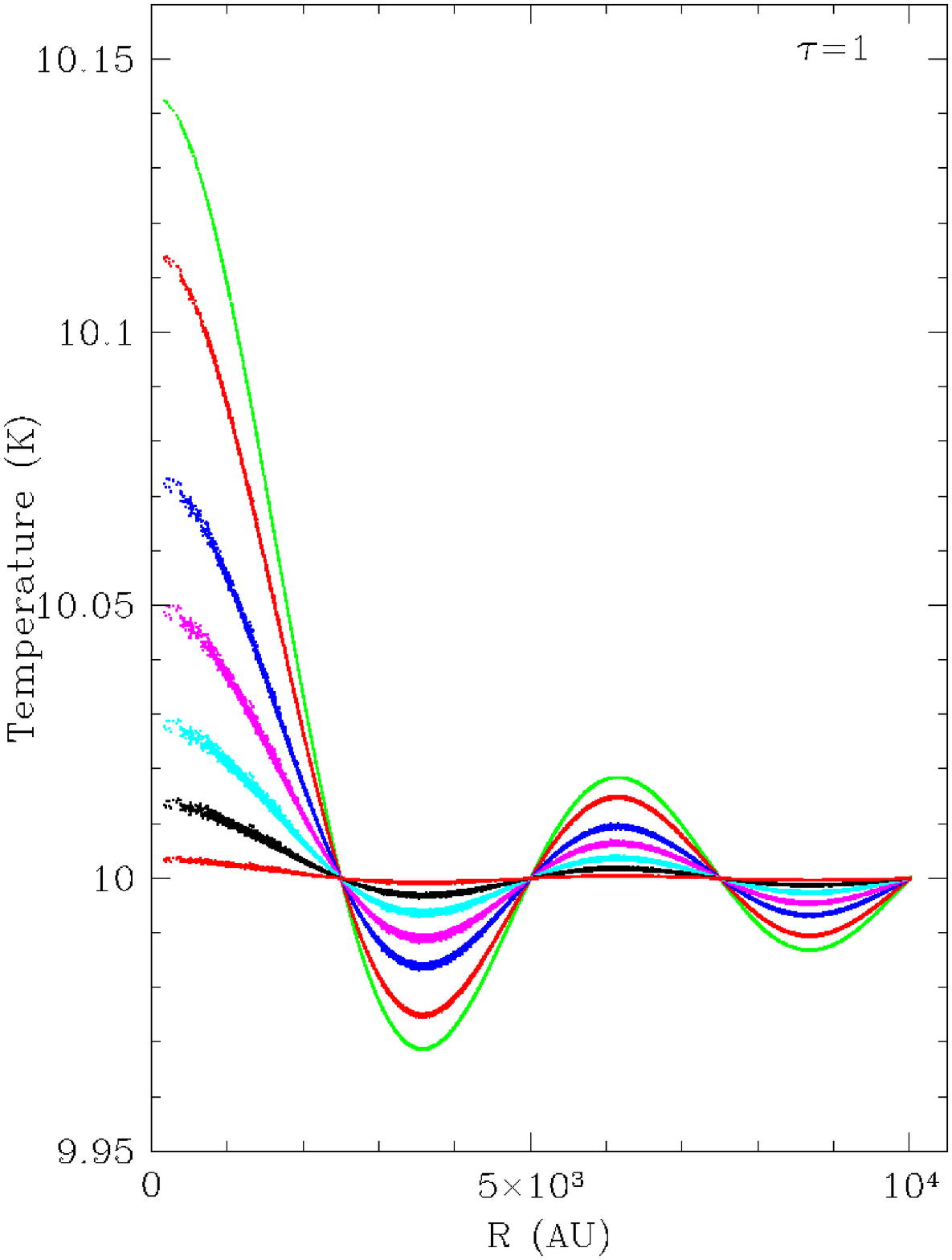}}
\caption{Same as Fig.~\ref{fig:relax.tau=0.1}, but for a sphere of optical depth $\tau=1$.}
\label{fig:relax.tau=1}
\end{figure}
\begin{figure}[!h]
\centering{\includegraphics[height=8cm]{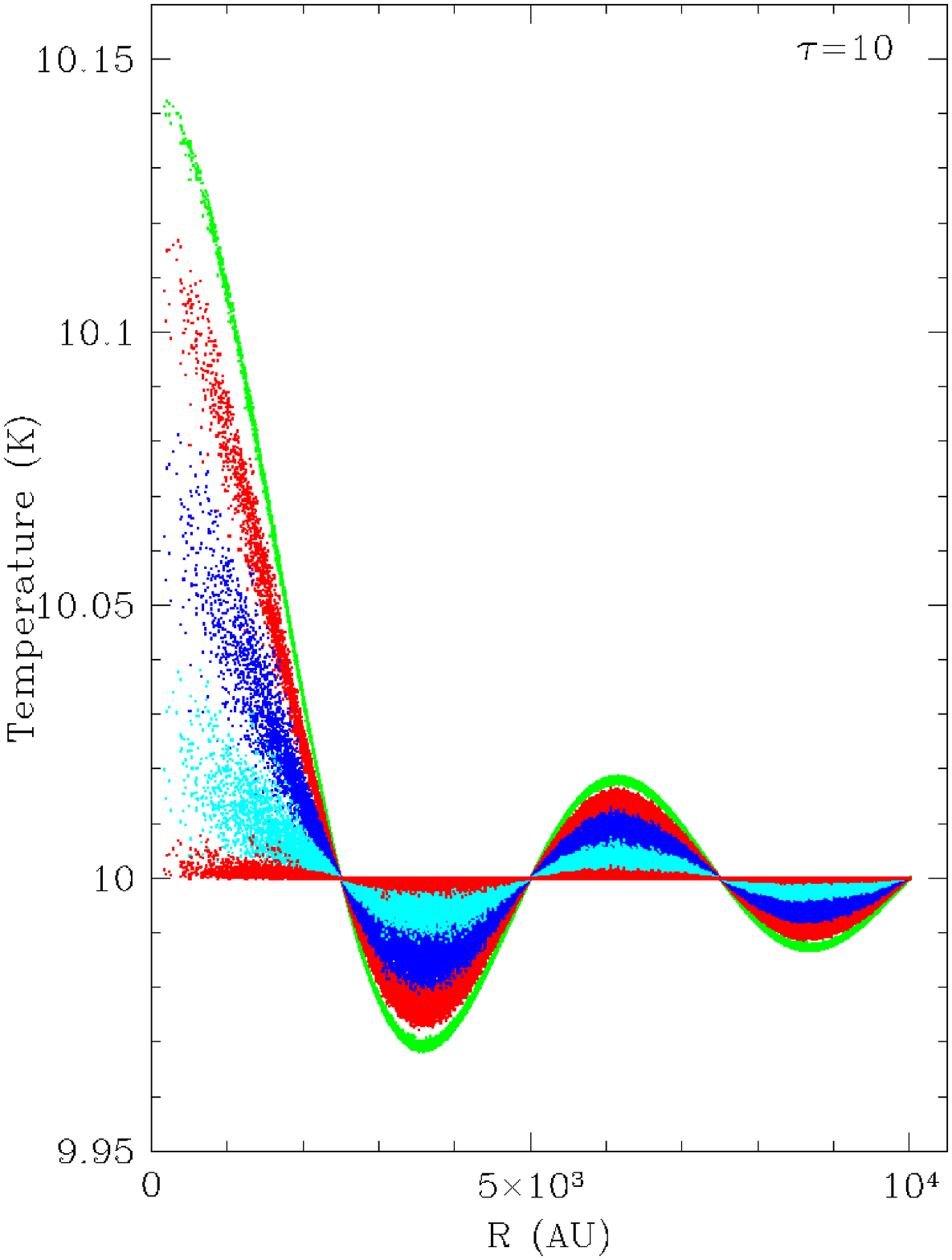}}
\caption{Same as Fig.~\ref{fig:relax.tau=0.1} but for a sphere of optical depth $\tau=10$.}
\label{fig:relax.tau=10}
%\end{figure}
%\begin{figure}
\centering{\includegraphics[height=8cm]{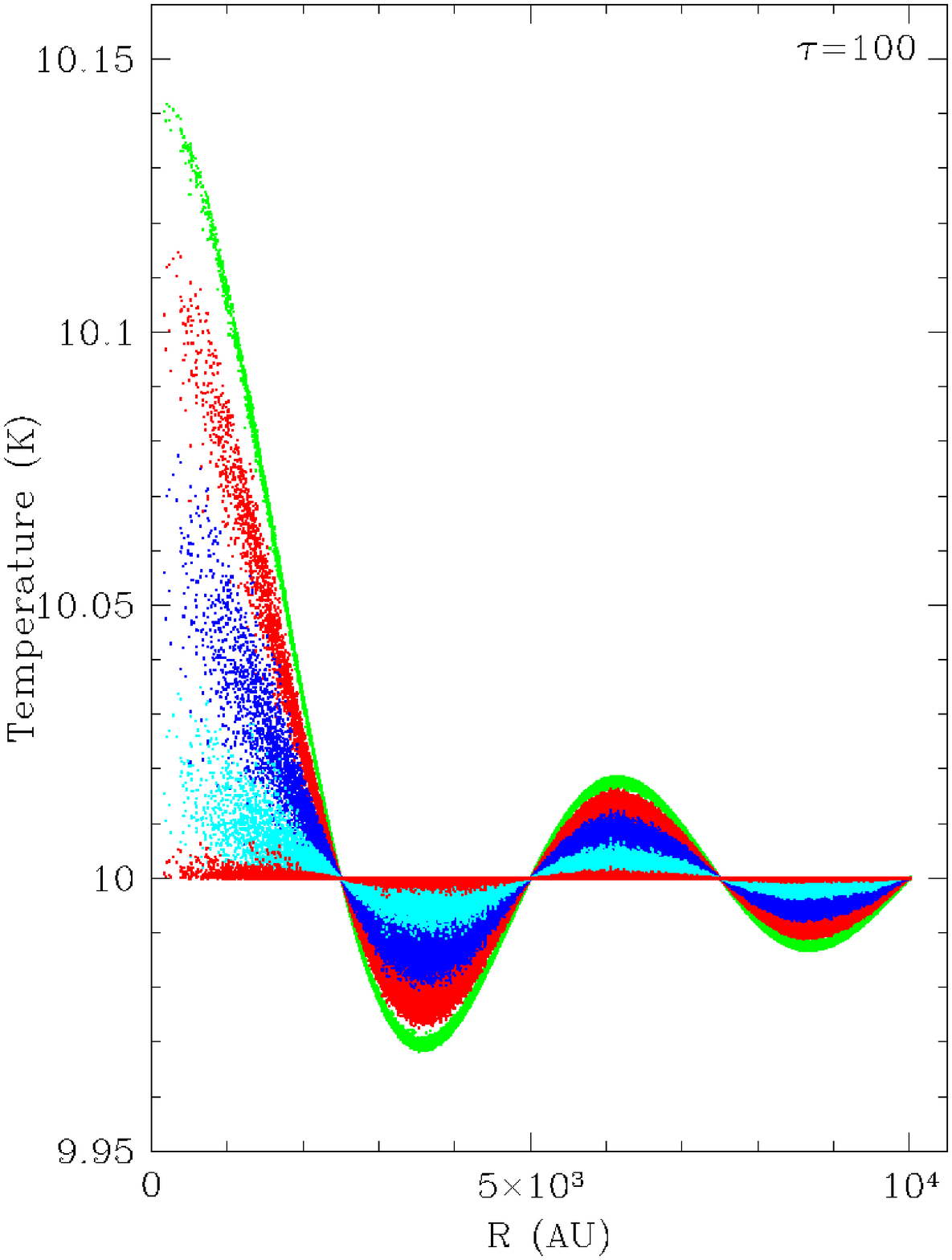}}
\caption{Same as Fig.~\ref{fig:relax.tau=0.1} but for a sphere of optical depth $\tau=100$.}
\label{fig:relax.tau=100}
\end{figure}
%%%%%%%%%%%%%%%%%%%%%%%%%%%%%%%%%%%%%%%%%%%%%%%%%%%%%%%%%%%%%%%

In Figs.~\ref{fig:relax.tau=0.1} to \ref{fig:relax.tau=100} we present the radial temperature profiles at different instants during the relaxation towards equilibrium for spheres having different optical depths, $\tau=0.1,\,1,\,10,\;{\rm and}\; 100$. The simulation results approximate well to Eqn. (\ref{EQN:RELAX}).

\begin{figure}
\centering{\includegraphics[height=8.5cm,angle=-90]{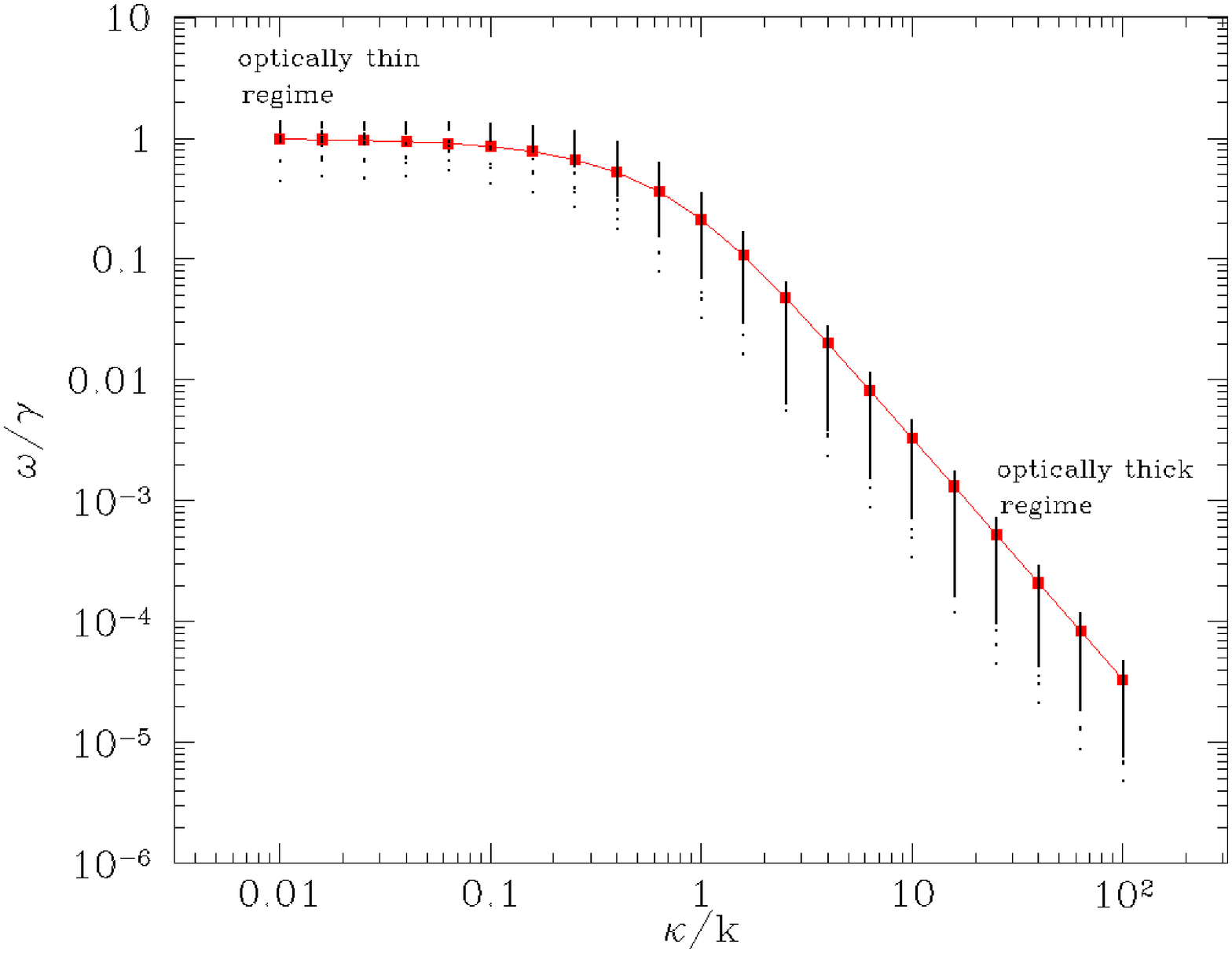}}
\caption{Dispersion relation for the thermal relaxation mode. The relaxation rates (in units of $\gamma$) of all SPH particles at seven different instants are plotted against the ratio $\kappa_{_{\rm O}}/k$ (dots that saturate to form lines). The theoretical values (calculated from Eq.~\ref{eq:rate}) are also plotted (red line and squares).}
\label{fig:relax.disp}
\end{figure}

The relaxation rates are also reproduced well. In Fig. \ref{fig:relax.disp} we present the dispersion relation, i.e. the relaxation rate $\omega(k)$ for different values of the ratio $\kappa_{_{\rm O}}/k$. We plot the relaxation rates for all the SPH particles that represent the uniform density sphere, calculated at seven different snapshots during the temperature relaxation (dots that saturate to form lines). The theoretical values (calculated from Eq.~\ref{eq:rate}) are also plotted (red line and squares).

%%%%%%%%%%%%%%%%%%%%%%%%%%%%%%%%%%%%%%%%%%%%%%%%%%%%%%%%%%%%%% FIGURE
%%%%%%%%%%%%%%%%%%%%%%%%%%%%%%%%%%%%%%%%%%%%%%%%%%%%%%%%%%%%%%%%%%%%%

%%%%%%%%%%%%%%%%%%%%%%%% SECTION
\section{Summary}\label{sec:sum}
%%%%%%%%%%%%%%%%%%%%%%%%%%%%%%%%

We have developed a new method to treat the influence of radiative transfer on the energy equation in SPH simulations of star formation. The method uses the density, temperature and gravitational potential of each particle to make an educated estimate of the mean optical depth which regulates its heating and cooling. It can treat both the optically thin and optically thick regimes.

The energy equation takes account of heating by compression or cooling by expansion (i.e. $P\,\!dV$ work); viscous dissipation; external irradiation; and radiative cooling. In situations where the thermal timescale is much longer than the dynamical timescale, the resulting thermal inertia effects are captured properly. Conversely, where the thermal timescale is much shorter than the dynamical timescale, we avoid very short timesteps, essentially by assuming thermal equilibrium. 

The equation of state and the internal energy take account of (i) the rotational and vibrational degrees of freedom of H$_2$, and (ii) the different chemical states of hydrogen and helium (cf. Black \& Bodenheimer 1975; Boley et al. 2007)

A simple parametrisation of the frequency averaged opacity is used (Bell \& Lin 1994), which reproduces the basic features of more sophisticated opacity models (e.g. Preibisch et al. 1993; Alexander \& Ferguson 1994). This parametrisation accounts for the effects of dust sublimation, molecules, H$^{-}$, free-free transitions, and electron scattering

To test the SPH-RT method we have examined the collapse of a $1\,{\rm M}_\odot$ molecular cloud of initially uniform density and uniform temperature. The collapse proceeds almost isothermally until the density in the centre rises above $\rho\sim 10^{-13}\ {\rm g cm}^{-3}$. Then the temperature rises rapidly, the thermal pressure decelerates the collapse, and the first core is formed. The first core grows in mass, and contracts quasistatically. When its temperature reaches $T\sim 2,000\,{\rm K}$, the H$_2$ starts dissociating and the second collapse starts, resulting in the formation of the second core, i.e. the protostar. Our method reproduces well the results of the detailed simulation of Masunaga \& Inutsuka (2000): the first and the second cores form at similar densities, having similar sizes, and at similar times after the start of the collapse.

We have also performed the Boss \& Myhill (1992) and Boss \& Bodenheimer (1979) tests, and obtained results very similar to those of Whitehouse \& Bate (2006). Finally, we have performed the thermal relaxation test of Masunaga et al. (1998). The geometries treated in this paper establish the fidelity of the method in treating both spherical and flattened geometries. Furthermore, the method also performs well on the Hubeny (1990) test, which deals with equilibrium discs, and hence it can also be applied to disc simulations. We will discuss the Hubeny (1990) test, and applications of this method to discs 
in a forthcoming paper (Stamatellos \& Whitworth, in preparation).

The new SPH-RT method performs very well, and most importantly it is very efficient. The computational time is almost the same as (only $\sim 3\,\%$ longer than) an SPH simulation using a barotropic equation of state. The method  is inherently three dimensional, and so it can be used to treat a variety of astrophysical systems, where the radiative processes and thermal inertia effects are important. We will report on applications of the method in future publications.

%%%%%%%%%%%%%%%%%%%%%%%%
\begin{acknowledgements}
We would like to thank the referee, S. Inutsuka, for his suggestions which helped to improve the original manuscript. The computations reported here were performed using the UK Astrophysical Fluids Facility (UKAFF). We thank S. Whitehouse and M. Bate for kindly providing data from Whitehouse \& Bate (2006), C. Clarke, M. Bate, I. Bonnell and
R. W\"{u}nsch for useful discussions, and D. Price for the use of {\sc splash/supersphplot}. We also acknowledge support by PPARC grant PPA/G/O/2002/00497. 
\end{acknowledgements}

%%%%%%%%%%%%%%%%%%%%%%%%%

\end{document}